\documentclass[pdftex,twocolumn]{aastex63}

\usepackage{apjfonts,graphicx,graphics}

\usepackage{hyperref}
\usepackage{graphicx}
\usepackage{mathptmx}
\usepackage{amsmath}
\usepackage{wasysym}
\usepackage{times}
\usepackage{epsfig}
\usepackage{ulem}
\usepackage{epstopdf}
\usepackage{appendix}
\usepackage{units}
\usepackage{import}
\newbox{\bigpicturebox}

\usepackage[capitalize]{cleveref}
\usepackage{CJKutf8}

\newcommand{\chinesenameXiaoweiduan}{{\begin{CJK}{UTF8}{gbsn}(段晓苇)\end{CJK}}}

\newcommand{\chinesenameXiaodianchen}{{\begin{CJK}{UTF8}{gbsn}(陈孝钿)\end{CJK}}}

\newcommand{\chinesenameLicaideng}{{\begin{CJK}{UTF8}{gbsn}(邓李才)\end{CJK}}}

\newcommand{\chinesenameFanyang}{{\begin{CJK}{UTF8}{gbsn}(杨帆)\end{CJK}}}

\newcommand{\chinesenameChaoliu}{{\begin{CJK}{UTF8}{gbsn}(刘超)\end{CJK}}}

\newcommand{\chinesenameHuaweizhang}{{\begin{CJK}{UTF8}{gbsn}(张华伟)\end{CJK}}}

\newcommand{\chinesenameWeijiasun}{{\begin{CJK}{UTF8}{gbsn}(孙唯佳)\end{CJK}}}

\newcommand{\m}{\rm\; m}

\newcommand{\s}{\rm\; s}

\newcommand{\K}{\rm\; K}
\newcommand{\g}{\rm\; g}
\newcommand{\W}{\rm\; W}
\begin{document}

\title{Blueshifted hydrogen emission and shock wave of RR Lyrae variables in
SDSS and LAMOST}

\author[0000-0002-6573-6719]{Xiao-Wei Duan \chinesenameXiaoweiduan}

\affiliation{Department of Astronomy, Peking University, Yi He Yuan Road 5, Hai
Dian District, Beijing 100871, China;
\href{mailto:duanxw@pku.edu.cn}{duanxw@pku.edu.cn}}
\affiliation{Kavli Institute for Astronomy \& Astrophysics, Peking University,
Yi He Yuan Road 5, Hai Dian District, Beijing 100871, China}

\author[0000-0001-7084-0484]{Xiao-Dian Chen \chinesenameXiaodianchen}
\affiliation{CAS Key Laboratory of Optical Astronomy, National Astronomical
Observatories, Chinese Academy of Sciences, Beijing 100101, China;
\href{mailto:licai@bao.ac.cn}{licai@bao.ac.cn}}
\affiliation{School of Astronomy and Space Science, University of the Chinese
Academy of Sciences, Huairou 101408, China}
\affiliation{Department of Astronomy, China West Normal University, Nanchong
637009, China}

\author[0000-0002-3279-0233]{Wei-Jia Sun \chinesenameWeijiasun}
\affiliation{Department of Astronomy, Peking University, Yi He Yuan Road 5, Hai
Dian District, Beijing 100871, China;
\href{mailto:duanxw@pku.edu.cn}{duanxw@pku.edu.cn}}
\affiliation{Kavli Institute for Astronomy \& Astrophysics, Peking University,
Yi He Yuan Road 5, Hai Dian District, Beijing 100871, China}

\author[0000-0001-9073-9914]{Li-Cai Deng \chinesenameLicaideng}
\affiliation{CAS Key Laboratory of Optical Astronomy, National Astronomical
Observatories, Chinese Academy of Sciences, Beijing 100101, China;
\href{mailto:licai@bao.ac.cn}{licai@bao.ac.cn}}
\affiliation{School of Astronomy and Space Science, University of the Chinese
Academy of Sciences, Huairou 101408, China}
\affiliation{Department of Astronomy, Peking University, Yi He Yuan Road 5, Hai
Dian District, Beijing 100871, China;
\href{mailto:duanxw@pku.edu.cn}{duanxw@pku.edu.cn}}
\affiliation{Department of Astronomy, China West Normal University, Nanchong
637009, China}

\author[0000-0002-7727-1699]{Hua-Wei Zhang \chinesenameHuaweizhang}
\affiliation{Department of Astronomy, Peking University, Yi He Yuan Road 5, Hai
Dian District, Beijing 100871, China;
\href{mailto:duanxw@pku.edu.cn}{duanxw@pku.edu.cn}}
\affiliation{Kavli Institute for Astronomy \& Astrophysics, Peking University,
Yi He Yuan Road 5, Hai Dian District, Beijing 100871, China}

\author[0000-0002-1450-9727]{Fan Yang \chinesenameFanyang}
\affiliation{CAS Key Laboratory of Optical Astronomy, National Astronomical
Observatories, Chinese Academy of Sciences, Beijing 100101, China;
\href{mailto:licai@bao.ac.cn}{licai@bao.ac.cn}}

\author[0000-0002-1802-6917]{Chao Liu \chinesenameChaoliu}
\affiliation{School of Astronomy and Space Science, University of the Chinese
Academy of Sciences, Huairou 101408, China}
\affiliation{CAS Key Laboratory of Space Astronomy and Technology, National
Astronomical Observatories, Chinese Academy of Sciences, Beijing 100101, China}

\begin{abstract}
Hydrogen emissions of RR Lyrae variables are the imprints of shock waves
traveling through their atmospheres. We develop a pattern recognition
algorithm, which is then applied to single-epoch spectra of SDSS and LAMOST.
These two spectroscopic surveys covered $\sim$ 10,000 photometrically confirmed RR Lyrae stars. We discovered in total 127 RR Lyrae stars with blueshifted Balmer
emission feature, including 103 fundamental mode (RRab), 20 first-overtone
(RRc), 3 double-mode (RRd), and 1 Blazhko type (temporary
classification for RR Lyrae stars with strong Blazhko modulation in Catalina
sky survey that cannot be characterized) RR Lyrae variable. This forms  the largest database to date of the properties of hydrogen emission in RR Lyrae
variables. Based on ZTF DR5, we carried out a detailed light-curve analysis for
the Blazhko type RR Lyrae star with hydrogen emission of long-term modulations.
We characterize the Blazhko type RR Lyrae star as an RRab and point out a
possible Blazhko period. Finally, we set up simulations on mock spectra to test the performance of our algorithm and on the real observational strategy to investigate the occurrence of the ``first apparition''.
\end{abstract}

%%%%%%%%%%%%%%%%%%%%%%%%%%%%%%%%%%%%%%%%%%%%%%%%%%%%%%%%%%%%%%%%

\section{Introduction \label{sec:intro}}

RR Lyrae stars are population II stars commonly found in globular clusters.
They are located at the intersection of the instability strip and the
horizontal branch (HB), from A- to F-type \citep{Smith1995,Sesar2012}. RR Lyrae
variables are classified, according to the number of oscillation modes, as
fundamental mode (RRab), the first overtone mode (RRc), or double-mode
(RRd) stars \citep{Soszy2011AcA....61....1S}. They obey the
$M_{V}-[\mathrm{Fe/H}]$ relation \citep{Muraveva2018MNRAS.481.1195M} and the
period-luminosity-metallicity relationship
\citep[PLZ,][]{Longmore1986MNRAS.220..279L,Catelan2004ApJS..154..633C} in the
infrared passbands, which makes them good ``standard candles'' for precise
distance determinations for star clusters and nearby galaxies
\citep{Bhardwaj2020JApA...41...23B}. Some of the RR Lyrae variables,
across all subtypes, show periodic amplitude and/or phase modulations, which is
known as ``Blazhko effect"
\citep{Bla1907AN....175..325B,Kolenberg2006A&A...459..577K}.

Despite the successful application of RR Lyrae stars as a step of distance
ladder, the detailed dynamical evolutions of the interior are unclear
and complicated. However, we can speculate the internal structure of RR
Lyrae variables by tracing the imprints of feature lines on spectra. Hydrogen
emission lines, Helium emission lines, line broadening and doubling phenomena,
neutral metallic line disappearance phenomena in RR Lyrae stars indicate shock
waves propagating through their atmospheres
\citep{Chadid2017ApJ...835..187C,Chadid2008A&A...491..537C}. The emissions are
produced when atoms in the gas de-excite after being excited by a shock wave
traveling into the atmosphere. The shock provides high enough energy to excite
neutral hydrogen from the second quantum state upwards. It compresses, heats,
and accelerates the gas. After that, the atoms release energy in the cooler
shock wake region and generate the emission \citep{Gillet2014A&A...565A..73G}.

Hydrogen emission lines of RR Lyrae stars are called ``apparitions''
\citep{Preston2011AJ....141....6P}. There are three famous ``apparitions'',
which got their sequence number based on the discovery time. \cite{Sanford1949}
firstly reported a hydrogen emission in RR Lyrae in 1949. The ``first
apparition'' is an intense blueshifted emission, which mostly presents during
$\phi\approx0.91$ (in RRab), just before the maximum luminosity
\citep{Gillet2019}. \cite{Duan2021ApJ...909...25D} reported the first detection
of blueshifted hydrogen emission (the ``first apparition'' at both H$\alpha$
and H$\beta$) in the first-overtone and double-mode RR Lyrae variables. Now it
is well accepted that the ``first apparition'' is generated from the hot
emitting layer behind an outward-moving shock front \citep{Schwarzschild1952}.

The ``second apparition'' is a weak hump at the blue side of the absorption
line wing, when $\phi\approx0.73$, discovered in 1988 \citep{Gillet1988}. It is
generated by the collision when layers of the upper atmosphere meet the
photospheric layers during the in-falling processes of the ballistic motion. It
occurs during the small bump around $\phi\approx0.72$, which was produced by
the ``early shock'' \citep{Hill1972,Gillet2014A&A...565A..73G}. Moreover, the
third one was discovered in 2011 \citep{Preston2011AJ....141....6P}, which is a
weak redshifted emission line that appears at around $\phi\approx0.30$. There
are two views on the origin of the ``third apparition''. One is that it is
possibly provided by a weakly supersonic and infalling shock wave of post
maximum (called $Sh_{PM}$), which is generated from superimposing compression
due to hydrogen recombination front and the accumulations of weak compression
waves \citep{Chadid2013}. The other opinion is that this emission is a P-Cygni
like profile, which is the sign that the envelope surrounding the stellar
surface is expanding, and the emission was produced when the shock wave is
detached from the photosphere. To confirm which one is closer to the fact,
further investigation is still needed.

Thanks to the rapid improvement of observing facilities, we can get an
ever-expanding volume of observed spectroscopic data of RR Lyrae variables. In
this work, we develop a pattern recognition searching algorithm and apply it to
a large database of single-epoch spectra, the ``first apparition''. We
find out spectra of RR Lyrae stars which show most prominent observational
characteristics of shock waves. We report hydrogen emissions in 103 RRab, 20
RRc, 3 RRd, and 1 Blazhko type RR Lyrae variables \footnote{Some of the
RR Lyrae stars with strong Blazhko modulation in Catalina sky survey can not be
accurately classified due to the significant modulation. They are temporarily
classified as ``Blazhko type RR Lyrae variable''. But they are not all of the
stars which show Blazhko effect.}. The detection of blueshifted hydrogen
emission in non-fundamental mode RR Lyrae variables has been discussed in our
previous work \citep{Duan2021ApJ...909...25D}. In this work, we present the
results of RRab variables and the Blazhko type RR Lyrae star (RR-Bl, the Blazhko type RR
Lyrae star with LAMOST spectra and hydrogen emission). The Blazhko type RR
Lyrae star is characterized as an RRab star using pre-whitening sequence
method. We build up the largest database of properties of the ``first
apparitions'' in RR Lyrae stars based on SDSS and LAMOST, to investigate shock
waves in RR Lyrae stars with a new view.

This paper is organized as follows. In Section \ref{section:observations}, we
describe the observations of both photometry and spectroscopy. The
searching methods are interpreted in Section \ref{section:methods}. We
display the results of the search in Section \ref{section:Rs}. Moreover, we
analyze the frequency components in RR-Bl, characterize it as an RRab star, and
provide a possible Blazhko period in Section \ref{section:Blazhko}. The
detection rate and mock spectra test are discussed in Section
\ref{section:detectionrateproblem}. Finally, the conclusions are covered in
Section~\ref{section:conclusion}.

\section{Observations}\label{section:observations}

\subsection{Photometric observations}

Both photometric and spectroscopic observations are needed to hunt for the
``first apparitions''. Light curves are used to identify RR Lyrae variables.
They can be collected from the Catalina Sky Surveys, the Wide-field Infrared
Survey Explorer (WISE), the All Sky Automated Survey for Supernovae (ASAS-SN)
and
  the Asteroid Terrestrial-impact Last Alert System (ATLAS).

The Catalina Sky Survey began in 2004 \citep{Drake2009ApJ...696..870D}. It uses
three telescopes to cover the sky ($-75^\circ<\delta<+65^\circ$), in order to
search for Near Earth Objects (NEOs) and Potential Hazardous Asteroids (PHAs).
The sub-surveys are specified as the Catalina Schmidt Survey (CSS), the Mount
Lemmon Survey (MLS) located in Tucson Arizona, and the Siding Spring Survey
(SSS) which is located in Siding Spring Australia. In Catalina survey, RR Lyrae
stars with strong Blazhko modulation are classified as ``Blazhko type'' other
than the basic types (RRab, RRc, and RRd) of RR Lyrae variables, because the
long-term modulation strongly influences the classification.
Lots of photometric observations and analyses of RR Lyrae stars have been
provided by related works
\citep{Drake2013,Drake2013b,Torrealba2015,Drake2014,Drake2017}.

Moreover, the Wide-field Infrared Survey Explorer \citep[WISE,][] {chen2018} is
a 40 cm space telescope, designed to implement an all-sky survey in 4 MIR
bands, including $W$1 (3.35$\mu$m), $W$2 (4.60$\mu$m), $W$3 (11.56$\mu$m) and
$W$4 (22.09$\mu$m) \citep{Wright2010}. \cite{chen2018} has compiled the first
all-sky mid-infrared variable-star catalog based on WISE five-year survey data,
providing 1231 RR Lyrae stars newly identified.

The All Sky Automated Survey for Supernovae
\citep[ASAS-SN,][]{Shappee2014ApJ...788...48S,Kochanek2017PASP..129j4502K} is
currently consisting of 24 telescopes around the globe. It is now automatically
surveying the entire visible sky every night down to $\sim 18$ mag.
Catalog from ASAS-SN \citep{Jayasinghe2019MNRAS.486.1907J} provides $\sim
412,000$ variables including $\sim 8000$ periodic pulsating stars.

The Asteroid Terrestrial-impact Last Alert System
\citep[ATLAS,][]{Heinze2018AJ....156..241H} scans most of the sky
every night to search dangerous asteroids, which is also used to search for
photometric variability \citep{Tonry2018}. We also get light curves from the
Zwicky Transient Facility \citep[ZTF,][]{Bellm2019PASP..131a8002B,Masci2019PASP..131a8003M,Chen2020} and
parameters from {Gaia} DR2 \citep{Clementini2019A&A...622A..60C} for our
selected stars if available.

\subsection{Spectroscopic observations}

Our spectroscopic data was collected from the Sloan Digital Sky Survey (SDSS)
and the Large Sky Area Multi-Object Fiber Spectroscopic Telescope survey
(LAMOST). In this work, we adopt low-resolution and single epoch spectra. The
periods of RR Lyrae stars are very short. Under the circumstances, the emission
features can be overwhelmed by long-time exposures and the processes of
combining spectra to get a higher signal-to-noise ratio. Therefore, we use
low-resolution and single epoch spectra to capture these features which will
fast wear away otherwise.

The Sloan Digital Sky Survey \citep[SDSS,][]{Eisenstein2011AJ....142...72E} is
one of the most ambitious surveys in the history of astronomy and enjoys
enormous influence. It saw its first light in 1998 and entered routine
observations in 2000. Using the Sloan Foundation 2.5m optical telescope (Apache
Point Observatory, New Mexico), the SDSS serves images and spectra of the
Northern sky, and currently extends to the Southern Sky using the 2.5m Du Pont
optical telescope (Las Campanas Observatory, Chile). SDSS-III is a massive spectroscopic survey focusing on the distant universe, the Milky
Way Galaxy, and extrasolar planetary systems
\citep{Eisenstein2011AJ....142...72E}. As for the SDSS dataset, some of the
spectra come from the BOSS survey \citep{Dawson2013AJ....145...10D}. Spectra
from SDSS cover wavelength from 3800--9200 \AA $ $ and those from BOSS have
wavelength coverage of 3650--10400 \AA. They achieve $R \sim 1500$ at 3800 \AA
$ $ and $R \sim 2500$ at 9000 \AA. In this work, we use SDSS DR9
because we utilize parameter from the SEGUE Stellar Parameter Pipeline
\citep[SSPP,][]{Lee2008AJ....136.2022L}--DR9.

Large sky Area Multi-Object fiber Spectroscopic Telescope
\citep[LAMOST,][]{Deng2012RAA....12..735D,Zhao2012,Cui2012} is a Chinese
national scientific research facility. This survey is consisting of two major
parts: the LAMOST ExtraGAlactic Survey (LEGAS) and the LAMOST Experiment for
Galactic Understanding and Exploration (LEGUE). It sets up a spectroscopic
survey of millions of objects in much of the northern sky and has great
potential to survey a large volume of space efficiently. We apply for
low-resolution and single exposure spectra in LAMOST DR6 ($R \sim 1800$).
Each plate was observed more than 3 times for SDSS and 1-3 times for
LAMOST survey. As for SDSS, the exposure time is 10-40 minutes each. And as for
LAMOST,
that
is
10-30 minutes each. It provides the finest time resolution for RR Lyrae
targets. 

From this large database, we can screen out spectra of RR Lyrae stars.
We cross-match the catalog of RR Lyrae stars and the dataset of spectra with
topcat \citep{Taylor2005ASPC..347...29T} and apply for single-epoch spectra
from LAMOST. We get single-epoch spectra of 3526 RR Lyrae stars from SDSS DR9
and 5571 RR Lyrae stars from LAMOST DR1--6. The Location of RR Lyrae stars of
this sample are shown in Figure~\ref{fig:spatialdistribution}.

\section{Methods}\label{section:methods}

Based on the profiles of the ``first apparitions'', we adapt two methods to
search for this kind of feature. In this section, we will explain the
technical details of the pipeline we use to catch the blueshifted hydrogen
emission lines of RR Lyrae stars in a large sky survey database.

\subsection{1D pattern recognition method}

The normal practice to search for a target feature is to check the profiles of
spectra visually. Due to rapid growth of datasets in size and complexity, data
science is being introduced into astronomical studies
\citep{Pesenson2010,Baron2019}. Pattern recognition means automated recognition
of patterns and regularities in data. It has become a useful tool in fetching
specific features and dealing with classification \citep{Djorgovski2006}.

Given that the profile we want is quite simple and clear, we adopt a 1D pattern
recognition method with hand-crafted rules. We can trace the trend of the
profile as an intense emission Gaussian-like profile at the shoulder of the
blue wing of a Gaussian-like absorption line wing. Both emission and absorption
signatures should be more significant than $2\sigma$.

We choose windows as 6540--6590 \AA $ $ for H$\alpha$, 4840--4880 \AA $ $ for
H$\beta$. We presuppose that the minimum values in the selected windows
indicate the location of Balmer absorption profiles. The adopted results at
least show clear target patterns at H$\alpha$ and H$\beta$ simultaneously. Both
emission and absorption components should contain at least two observational
points. We apply the pipeline to our collected spectra sample, with visual
checks to avoid a false negative.

\subsection{Cross-correlation method}

Given that the appearance of the emission lines introduces sudden change on the
spectra, the cross-correlation method is also useful for selecting blueshifted
hydrogen emissions \citep{Yang2014}. If there is not a ``first apparition'',
there should only be one wide absorption line of H$\alpha$ on the spectrum. On
the contrary, there would be one additional emission line, which introduces a
big difference in the shapes of two spectra. So if the difference between
radial velocities is strangely large, there are possibly the ``first
apparitions''. 

\begin{figure*}[tb]
\begin{center}
\begin{tabular}{cc}
\includegraphics*[width=90.3mm,height=6cm]{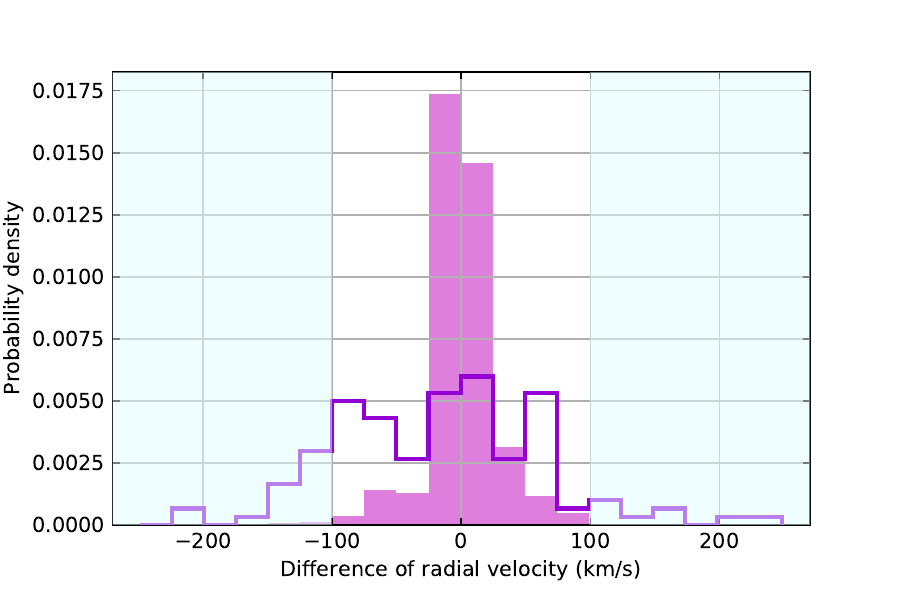}&
\includegraphics*[width=90.3mm,height=6cm]{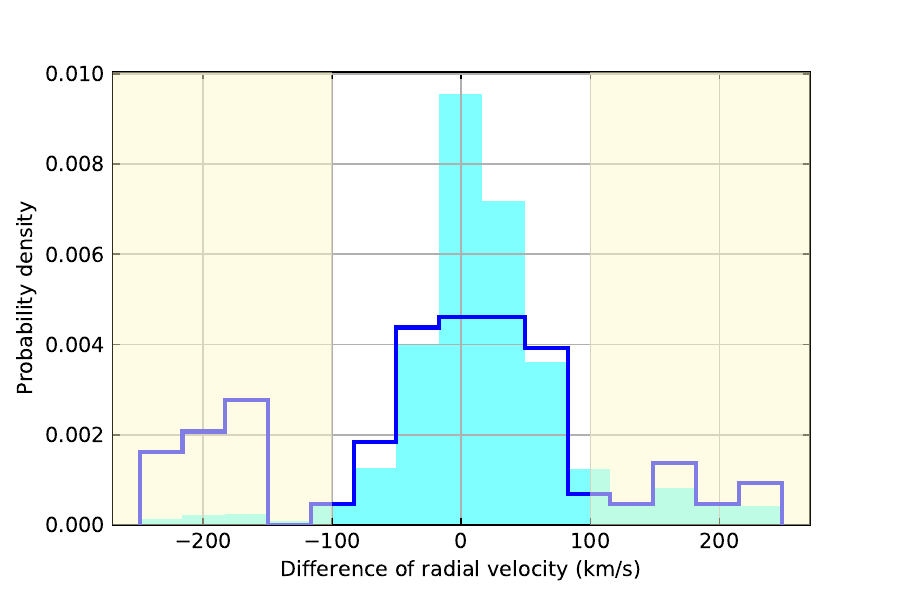}
\end{tabular}
\caption{\label{fig:dRV} Distribution of the differences of radial velocities
for SDSS (left) and LAMOST (right) respectively. The dataset was cleaned by
S/N$\geq$15 and $3\sigma$ criteria. The filled histogram denotes the
whole sample. The hollow histogram represents the distribution of the
differences of radial velocities for the selected stars in this work. As for
the cross-correlation method, we try to seek for the ``first apparition'' when
$|\Delta\mathrm{RV}|$ > 100 km/s.}
\end{center}
\end{figure*}

The cross-correlation method implemented here works as follows: We use the
first observed spectrum in a pair as the template, while the second one is
shifted by 1 km/s steps between -250 km/s to 250 km/s. The shifted template is
linearly interpolated at the points of the wavelength of the observational
spectrum to calculate the cross-correlation function. It shows the difference
of radial velocities ($\Delta$RV) between two spectra when the
cross-correlation function reaches its maximum. Here we provide the results of
differences of radial velocities with H$\alpha$ line in Figure~\ref{fig:dRV}.

We locally normalize the spectra between $6540 < \lambda < 6590 \AA$ by the
fitting result of the continuum. The quantity $\sigma$ of the spectra is
calculated from RMS of the spectra from the first and last 5 points, within
which the spectra are continuum dominated. We adopted results with both spectra
satisfying S/N > 15, and the significance of the absorption line is over
$3\sigma$. The distribution of $\Delta$RV for both SDSS and LAMOST dataset are
visualized in Figure~\ref{fig:dRV}. We check for the blueshifted hydrogen
emission when $|\Delta\mathrm{RV}|$ > 100 km/s.

All of the selected results are produced by the 1D pattern recognition
pipeline. The cross-correlation method provides 13 RRab stars, 1 RRc star for
SDSS and 44 RRab stars, 3 RRc stars, 1 RRd star and 1 Blazhko type RR Lyrae
star for LAMOST. The distribution of $\Delta$RV for selected RR Lyrae stars are
visualized in Figure~\ref{fig:dRV}. The cross-correlation method does help but
not all of the spectra with emission show high $\Delta$RV, and not all spectra
with high $\Delta$RV show emission. So our work is mainly based on the 1D
pattern recognition pipeline. The cross-correlation method is used as a
supplement.

\section{Results}\label{section:Rs}

\begin{figure*}[tb]

\begin{center}
\begin{tabular}{c}

\includegraphics*[width=120mm,height=8cm]{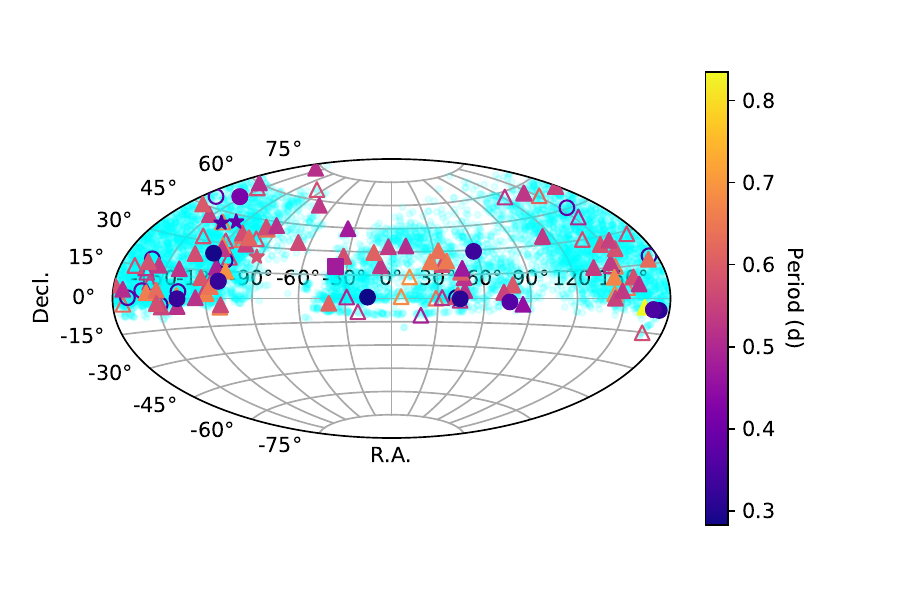}
\end{tabular}
\caption{\label{fig:spatialdistribution} 
Spatial distribution of selected targets with the ``first apparition'', with the color denoting the value of period. Corresponding relation:
hollow
triangle - RRab from SDSS, filled triangle - RRab from LAMOST, hollow circle -
RRc from SDSS, filled circle - RRc from LAMOST, star - RRc from LAMOST, square
- one Blazhko type star in LAMOST. Light blue points in the background denote
the whole sample which our survey is based on.}
\end{center}
\end{figure*}

As the result, we find 33 RRab stars and 10 RRc stars in SDSS, 70 RRab stars,
10 RRc stars, 3 RRd stars and 1 Blazhko type RR Lyrae star in LAMOST that show
the ``first apparition''. Visualization of the spatial distribution of the
selected stars is shown in Figure~\ref{fig:spatialdistribution}. The detailed
analyses of the results of RRc and RRd stars has been demonstrated in
\cite{Duan2021ApJ...909...25D}. In this paper, we present the measurements of
RRab and Blazhko type RR Lyrae variables and compare the properties between
different types of RR Lyrae stars.

The emission and absorption components are fitted with the $scale$ $width$
$versus$ $shape$ method for the hydrogen lines
\citep{Sersic1968adga.book.....S,Clewley2002MNRAS.337...87C}. We use two
$S\acute{e}rsic$ functions \citep{Xue2008,Yang2014,Duan2020CoBAO..67..181D} as:
\begin{eqnarray}\label{eq:Sersicprofile}
y =m - a e^{-(\frac{\left|{\lambda}-{\lambda_0}\right|}{b})^c},
\end{eqnarray}
to fit the profile. Uncertainties are provided by error propagation with the
covariance matrix and Monte Carlo method \citep{Andrae2010arXiv1009.2755A}.

\begin{figure*}[tb]
\begin{center}
\begin{tabular}{c}
\includegraphics*[width=180mm,height=21.6cm]{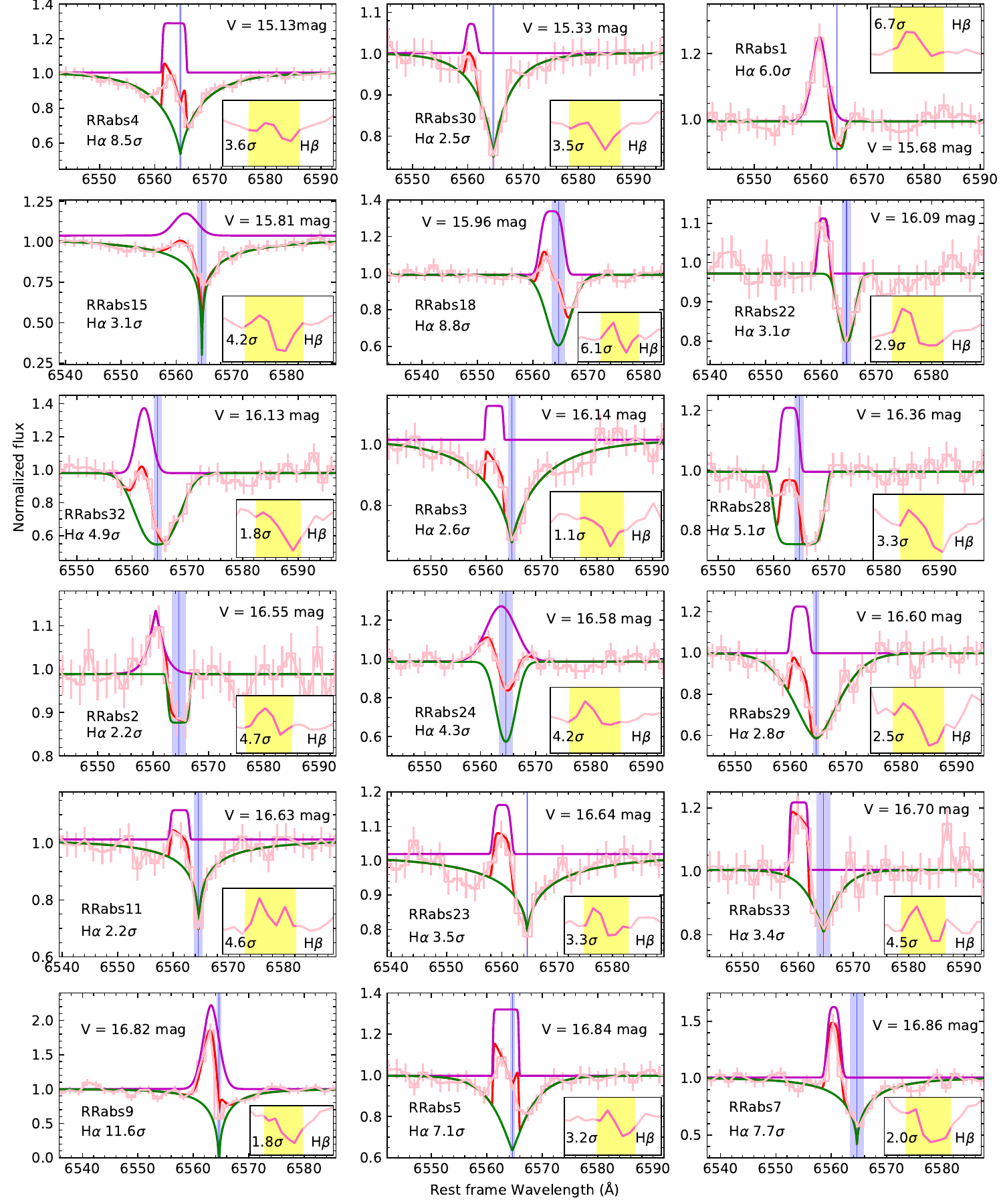}
\end{tabular}
\caption{\label{fig:SDSSexample} Visualization of examples of fitting results
of RRab that show hydrogen emission in SDSS. The wavelength axis is in the
stellar rest frame. The fitted blueshifted H$\alpha$ emission lines are shown
as pink profiles. Green profiles denote fitted absorption lines. Red profiles
represent the combination of fitted emission and absorption lines. Vertical
blue lines denote the H$\alpha$ line laboratory wavelength. The H$\beta$
profiles are shown in the subplots. The significances of the emission lines
denote the signal-to-noise ratio. The examples are ordered by $V$.}\end{center}
\end{figure*}

\begin{figure*}[tb]
\begin{center}
\begin{tabular}{c}
\includegraphics*[width=180mm,height=21.6cm]{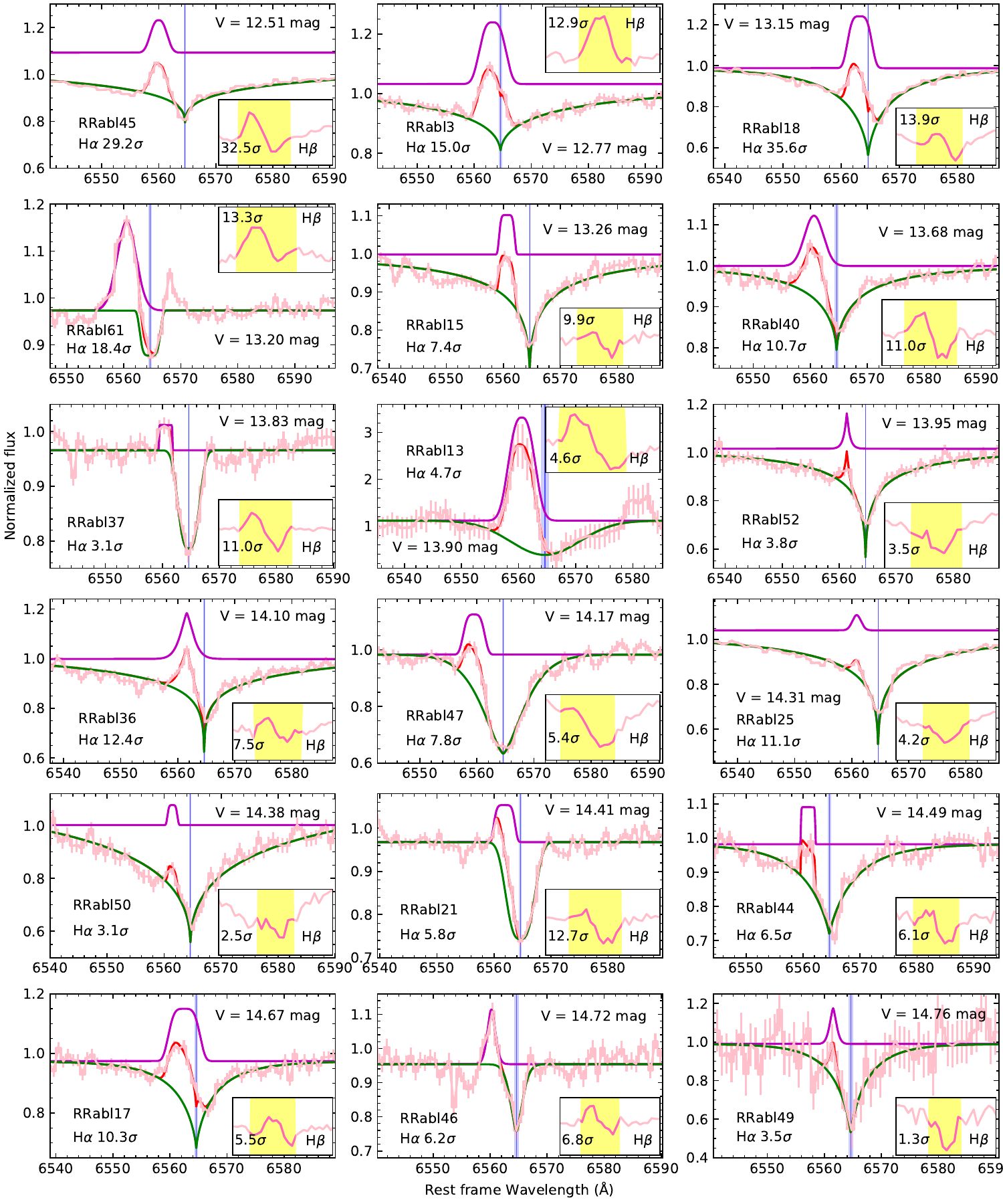}
\end{tabular}
\caption{\label{fig:LAMOSTexample}
As Figure~\ref{fig:SDSSexample}, but for selected RRab stars in
LAMOST.}\end{center}
\end{figure*}

\begin{figure*}[tb]
%\centering
\begin{center}
\begin{tabular}{c}
%\centering
\includegraphics*[width=120mm,height=8cm]{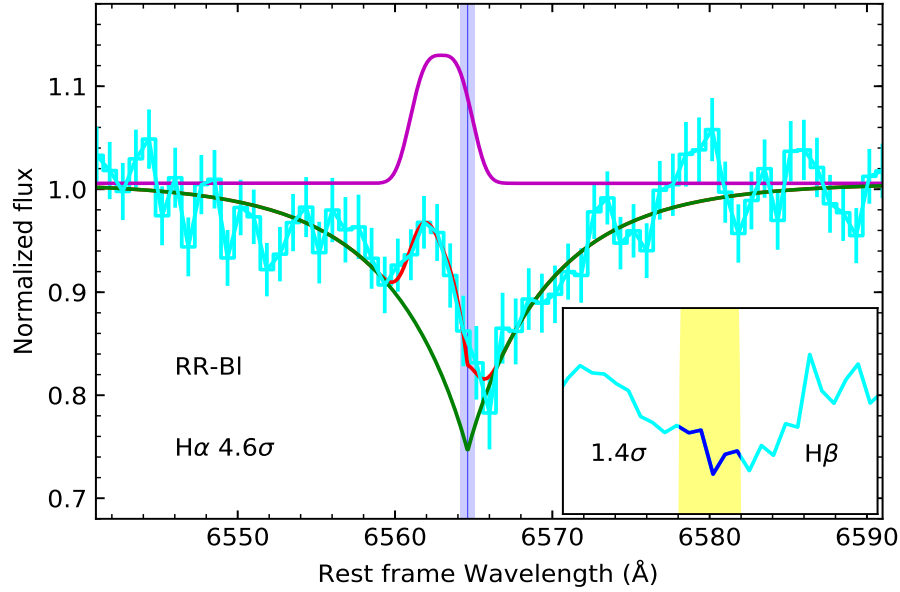}
\end{tabular}
\caption{\label{fig:LAMOSTBlaexample} As Figure~\ref{fig:SDSSexample}, but for
a selected Blazhko type RR Lyrae star in LAMOST.}
\end{center}
\end{figure*}

Figures~\ref{fig:SDSSexample} and \ref{fig:LAMOSTexample} show the fitting
examples of selected RRab stars in SDSS and LAMOST respectively, which display
the blueshifted emission in H$\alpha$. Figure~\ref{fig:LAMOSTBlaexample} shows
the fitting result of the Blazhko type RR Lyrae star with the ``first
apparition'' detected.
The wavelength axis is in the stellar rest frame. Flux is normalized by
continuum. The H$\alpha$ emissions are shown as pink profiles. H$\beta$
profiles are shown in the subplots. The significances of the blueshifted
emissions for both H$\alpha$ and H$\beta$ are also shown. The candidates are
adopted when the signals of H$\alpha$ is over $2\sigma$ and the signals of
H$\beta$ is over $1\sigma$.

The radial velocity of the blueshifted hydrogen emission in the stellar rest
frame is calculated as:
\begin{eqnarray}\label{eq:Vshock}
V_{\rm e1,\alpha} = c\frac{(\lambda_{\rm e1,\alpha}-\lambda_{\rm
ab})}{\lambda_{0}},
\end{eqnarray}
where $\lambda_{\rm e1,\alpha}$ denotes the wavelength corresponding to the
central wavelength of the emission line. $\lambda_{\rm ab}$ denotes the central
wavelength of the absorption line. $\lambda_{0}$ denotes the laboratory
wavelength.

We provide measurements of redshift $z_{\rm
e1,\alpha}$ and radial velocity in the stellar rest frame $V_{\rm e1,\alpha}$,
normalized flux Flux$_{\rm e1,\alpha}$ and full width at half maximum (FWHM) of
the emission and absorption FWHM$_{\rm e1,\alpha}$ in the stellar rest frame of
the blueshifted hydrogen emission.
Parameters of the selected RR Lyrae variables and measurements of the ``first
apparitions'' are summarized in Table~\ref{table:parameterSDSS1}, and
\ref{table:parameterLAMOST1}. RR Lyrae stars with SDSS spectra that
exhibit the first apparition are marked as RRabs (s as SDSS) and those with
LAMOST spectra and hydrogen emission as RRabl (l as LAMOST). The star ``RR-Bl''
denotes the Blazhko type RR Lyrae star with LAMOST spectra and hydrogen
emission (l as LAMOST). It was temporarily classified as a ``Blazhko type RR
Lyrae variable'' because it was not accurately classified due to the
significant modulation in the Catalina sky survey. In this work, RR-Bl is
characterized as an RRab star using pre-whitening sequence method in Section
\ref{section:Blazhko}.

\begin{table*}[tbp]
%\centering
\caption{\label{table:parameterSDSS1} Parameters of selected RRab stars from
SDSS.}
\begin{center}
\setlength{\tabcolsep}{1mm}{
\begin{tabular}{lccccccccccccccc}
%\hline
\hline
{Object}   & R.A.(J2000) & Decl.(J2000) & Period & $V$ & Amp & $z_{\rm
e1,\alpha}$ & $V_{\rm e1,\alpha}$ &  Flux$_{\rm e1,\alpha}$ & FWHM$_{\rm
e1,\alpha}$\\
$  $          & ${}^{\circ}$ & ${}^{\circ}$ & day & mag & mag        & $ $ &
km/s &   & $\rm \mathring{A}$   \\
\hline
RRabs1  & $6.20764$ & $+1.21571$ & $0.68042$ & $15.68$ & $1.00$ & $
-4.72E-4\pm1.86E-4$ & $-141\pm79$ & $0.25$ & $3.05$ \\
RRabs2  & $11.87371$ & $+13.91740$ & $0.69140$ & $16.55$ & $0.83$ & $
-6.35E-4\pm1.87E-4$ & $-190\pm79$ & $0.15$ & $2.48$ \\
RRabs3  & $19.11867$ & $-10.71342$ & $0.48310$ & $16.14$ & $1.08$ & $
-4.84E-4\pm1.88E-4$ & $-145\pm80$ & $0.11$ & $3.32$ \\
RRabs4  & $28.70903$ & $+0.25027$ & $0.63698$ & $15.13$ & $1.14$ & $
-1.68E-4\pm1.87E-4$ & $-50\pm79$ & $0.28$ & $4.30$ \\
RRabs5  & $32.71725$ & $+0.63203$ & $0.51342$ & $16.84$ & $1.17$ & $
-1.77E-4\pm1.87E-4$ & $-53\pm79$ & $0.32$ & $4.69$ \\
RRabs6  & $44.24281$ & $-1.15499$ & $0.51943$ & $17.79$ & $0.97$ & $
-5.76E-4\pm1.88E-4$ & $-173\pm80$ & $0.21^{*}$ & $1.63$ \\
RRabs7  & $122.43165$ & $+57.50441$ & $0.52138$ & $16.86$ & $1.10$ & $
-6.37E-4\pm1.85E-4$ & $-191\pm79$ & $0.62$ & $2.52$ \\
RRabs8  & $139.15997$ & $+30.14284$ & $0.57204$ & $17.13$ & $0.89$ & $
-2.75E-4\pm1.88E-4$ & $-82\pm80$ & $0.29$ & $3.58$ \\
RRabs9  & $144.23392$ & $+2.39681$ & $0.68102$ & $16.82$ & $0.13$ & $
-2.16E-4\pm9.45E-5$ & $-65\pm34$ & $1.22$ & $3.27$ \\
RRabs10  & $146.65068$ & $+15.99970$ & $0.52099$ & $17.73$ & $0.85$ & $
-1.92E-4\pm1.86E-4$ & $-57\pm79$ & $0.48$ & $4.69$ \\
RRabs11  & $149.11227$ & $+15.52289$ & $0.65209$ & $16.63$ & $0.94$ & $
-5.20E-4\pm1.86E-4$ & $-156\pm79$ & $0.10$ & $3.44$ \\
RRabs12  & $151.83224$ & $+53.47885$ & $0.61787$ & $17.10$ & $0.61$ & $
-3.41E-4\pm1.15E-4$ & $-102\pm49$ & $0.34$ & $5.91$ \\
RRabs13  & $152.83864$ & $+5.17696$ & $0.64363$ & $17.05$ & $0.93$ & $
-2.46E-4\pm1.85E-4$ & $-74\pm79$ & $0.24$ & $4.70$ \\
RRabs14  & $153.59429$ & $+40.33244$ & $0.49947$ & $17.16$ & $0.98$ & $
-4.67E-4\pm1.88E-4$ & $-140\pm80$ & $0.13$ & $2.38$ \\
RRabs15  & $167.19900$ & $-15.19530$ & $0.57329$ & $15.81$ & $0.76$ & $
-4.55E-4\pm9.61E-5$ & $-136\pm49$ & $0.14$ & $4.84$ \\
RRabs16  & $172.30798$ & $+28.80851$ & $0.56680$ & $17.72$ & $0.71$ & $
-2.95E-4\pm1.85E-4$ & $-89\pm79$ & $0.17$ & $3.11$ \\
RRabs17  & $186.58595$ & $-2.44612$ & $0.62669$ & $17.19$ & $0.67$ & $
-2.68E-4\pm6.97E-5$ & $-80\pm60$ & $0.49$ & $4.51^{*}$ \\
RRabs18  & $189.30895$ & $+14.44229$ & $0.57156$ & $15.96$ & $0.89$ & $
-1.81E-4\pm1.86E-4$ & $-54\pm79$ & $0.35$ & $4.37$ \\
RRabs19  & $197.03942$ & $+13.81974$ & $0.51911$ & $17.32$ & $1.03$ & $
-2.64E-4\pm1.88E-4$ & $-79\pm80$ & $0.45$ & $3.39$ \\
RRabs20  & $199.00117$ & $+11.91443$ & $0.52698$ & $17.54$ & $1.07$ & $
-7.74E-5\pm6.52E-5$ & $-23\pm34$ & $0.52$ & $4.52$ \\
RRabs21  & $202.86013$ & $+7.47637$ & $0.54216$ & $17.11$ & $1.15$ & $
-4.81E-4\pm1.00E-4$ & $-144\pm42$ & $0.18$ & $2.78$ \\
RRabs22  & $204.98954$ & $+58.16789$ & $0.57165$ & $16.09$ & $0.93$ & $
-6.40E-4\pm1.88E-4$ & $-192\pm80$ & $0.14$ & $2.13$ \\
RRabs23  & $220.46840$ & $+31.95186$ & $0.57671$ & $16.64$ & $0.91$ & $
-6.96E-4\pm1.87E-4$ & $-209\pm79$ & $0.14$ & $3.24$ \\
RRabs24  & $223.62869$ & $+40.07005$ & $0.66732$ & $16.58$ & $0.63$ & $
-1.32E-4\pm1.87E-4$ & $-40\pm79$ & $0.29$ & $6.25^{*}$ \\
RRabs25  & $238.83554$ & $+31.23341$ & $0.56251$ & $17.31$ & $0.91$ & $
-2.78E-4\pm1.87E-4$ & $-83\pm79$ & $0.31$ & $3.27$ \\
RRabs26  & $242.93547$ & $+10.72981$ & $0.55029$ & $17.31$ & $1.02$ & $
-5.83E-4\pm1.14E-4$ & $-175\pm65$ & $1.14^{*}$ & $0.12$ \\
RRabs$27_1$  & $247.42582$ & $+35.28750$ & $0.63104$ & $17.38$ & $0.92$ & $
-3.45E-4\pm1.88E-4$ & $-103\pm80$ & $0.25$ & $4.42$ \\
RRabs$27_2$  & $247.42582$ & $+35.28750$ & $0.63104$ & $17.38$ & $0.92$ & $
-3.82E-4\pm1.54E-4$ & $-115\pm62$ & $0.16$ & $4.30$ \\
RRabs28  & $248.63333$ & $+22.76128$ & $0.55024$ & $16.36$ & $0.96$ & $
-2.63E-4\pm1.86E-4$ & $-79\pm79$ & $0.21$ & $3.63$ \\
RRabs29  & $249.81254$ & $+32.90717$ & $0.63611$ & $16.60$ & $0.91$ & $
-4.49E-4\pm1.88E-4$ & $-135\pm80$ & $0.23$ & $3.42$ \\
RRabs30  & $259.17579$ & $+34.27990$ & $0.57687$ & $15.33$ & $0.75$ & $
-6.08E-4\pm1.87E-4$ & $-182\pm79$ & $0.07$ & $2.01$ \\
RRabs31  & $261.94862$ & $+65.60181$ & $0.57442$ & $17.32$ & $0.55$ & $
-3.66E-4\pm2.27E-5$ & $-110\pm56$ & $0.77$ & $3.13$ \\
RRabs32  & $331.06979$ & $+0.96542$ & $0.50032$ & $16.13$ & $0.71$ & $
-3.76E-4\pm4.50E-5$ & $-113\pm37$ & $0.39$ & $3.26$ \\
RRabs33  & $338.12842$ & $-8.48267$ & $0.52279$ & $16.70$ & $1.09$ & $
-6.80E-4\pm1.87E-4$ & $-204\pm79$ & $0.21$ & $3.38$ \\
\hline
\end{tabular}}
\end{center}
\tablecomments{\\\hspace{\textwidth}
1. Period, $V$ and Amp (Amplitude) are produced by Catalina Sky
Survey.\\\hspace{\textwidth}
2. $z_{\rm e1,\alpha}$ represents the redshift of the emission component of the
``first apparition'' in the stellar rest frame.\\\hspace{\textwidth}
3. $V_{\rm e1,\alpha}$ represents the radial velocity of the emission component
of the ``first apparition'' in the stellar rest frame.\\\hspace{\textwidth}
4. Flux$_{\rm e1,\alpha}$ indicates the normalized flux of the emission.
\\\hspace{\textwidth}
5. FWHM$_{\rm e1,\alpha}$ indicates full width at half maximum of the emission.
\\\hspace{\textwidth}
6. The names like RRabs$27_1$ mean that there are not only one spectrum of one
star. \\\hspace{\textwidth}
7. The measurements with $^*$ denote low-quality results. \\\hspace{\textwidth}
8. The stars are ordered by R.A.(J2000).}
\end{table*}

\begin{table*}[tbp]
%\centering
\caption{\label{table:parameterLAMOST1} Parameters of selected RRab and Blazhko
type stars from LAMOST.}
\begin{center}
\setlength{\tabcolsep}{1mm}{
\begin{tabular}{lccccccccccccccc}
\hline
{Object}   & R.A.(J2000) & Decl.(J2000) & Period & $V$ & Amp & $z_{\rm
e1,\alpha}$ & $V_{\rm e1,\alpha}$ &  Flux$_{\rm e1,\alpha}$ & FWHM$_{\rm
e1,\alpha}$\\
$  $          & ${}^{\circ}$ & ${}^{\circ}$ & day & mag & mag        & $ $ &
km/s &   & $\rm \mathring{A}$   \\
\hline
RRabl1  & $26.68188$ & $+23.58775$ & $0.65160$ & $17.01$ & $0.90$ & $
-5.04E-4\pm9.89E-5$ & $-151\pm42$ & $0.14$ & $1.62$ \\
RRabl2  & $33.40705$ & $+30.47452$ & $0.63676$ & $15.73$ & $0.57$ & $
-5.33E-4\pm3.86E-5$ & $-160\pm17$ & $0.14$ & $1.92$ \\
RRabl3  & $34.39211$ & $+21.71650$ & $0.56668$ & $12.77$ & $0.67$ & $
-2.11E-4\pm1.51E-5$ & $-63\pm16$ & $0.21$ & $4.84$ \\
RRabl4  & $37.83211$ & $+23.74759$ & $0.64989$ & $16.90$ & $0.39$ & $
-4.69E-4\pm9.87E-5$ & $-141\pm42$ & $0.30$ & $1.14$ \\
RRabl$5_1$  & $42.25593$ & $+1.34632$ & $0.57374$ & $15.96$ & $0.93$ & $
-6.09E-4\pm9.87E-5$ & $-183\pm42$ & $0.07$ & $3.08$ \\
RRabl$5_2$  & $42.25593$ & $+1.34632$ & $0.57374$ & $15.96$ & $0.93$ & $
-3.71E-4\pm4.08E-5$ & $-111\pm16$ & $0.12$ & $2.11$ \\
RRabl6  & $42.86087$ & $+2.45366$ & $0.61441$ & $15.77$ & $0.85$ & $
-7.48E-4\pm4.53E-5$ & $-224\pm16$ & $0.12$ & $2.91$ \\
RRabl7  & $45.98163$ & $+16.53199$ & $0.59756$ & $17.02$ & $0.80$ & $
-6.27E-4\pm9.93E-5$ & $-188\pm42$ & $0.12^{*}$ & $2.26$ \\
RRabl8  & $47.38816$ & $+18.92607$ & $0.47025$ & $17.12$ & $1.18$ & $
-4.77E-4\pm5.94E-5$ & $-143\pm20$ & $0.19$ & $2.07$ \\
RRabl9  & $47.46535$ & $+5.02683$ & $0.51977$ & $17.69$ & $0.94$ & $
-5.86E-4\pm4.94E-5$ & $-176\pm16$ & $0.19$ & $1.98$ \\
RRabl10  & $47.61406$ & $+13.10191$ & $0.52153$ & $16.90$ & $0.96$ & $
-5.63E-4\pm9.89E-5$ & $-169\pm42$ & $0.14$ & $0.86$ \\
RRabl$11_1$  & $72.68140$ & $+3.79305$ & $0.55219$ & $16.11$ & $0.77$ & $
-2.36E-4\pm2.62E-5$ & $-71\pm21$ & $0.30$ & $4.65$ \\
RRabl$11_2$  & $72.68140$ & $+3.79305$ & $0.55219$ & $16.11$ & $0.77$ & $
-2.77E-4\pm8.91E-5$ & $-83\pm40$ & $0.44$ & $3.76$ \\
RRabl$12_1$  & $78.84653$ & $+8.10691$ & $0.59066$ & $16.62$ & $0.63$ & $
-6.09E-4\pm4.73E-5$ & $-183\pm31$ & $0.51$ & $3.85$ \\
RRabl$12_2$  & $78.84653$ & $+8.10691$ & $0.59066$ & $16.62$ & $0.63$ & $
-6.15E-4\pm4.99E-5$ & $-184\pm33$ & $0.45$ & $3.63$ \\
RRabl13  & $84.93823$ & $-3.42164$ & $0.45415$ & $13.90$ & $1.07$ & $
-6.20E-4\pm2.21E-5$ & $-186\pm31$ & $2.20$ & $5.00$ \\
RRabl14  & $113.32693$ & $+34.50516$ & $0.53281$ & $16.64$ & $0.88$ & $
-4.31E-4\pm3.47E-5$ & $-129\pm31$ & $0.25$ & $2.38$ \\
RRabl15  & $134.54286$ & $+15.80513$ & $0.54316$ & $13.26$ & $0.76$ & $
-6.08E-4\pm9.87E-5$ & $-182\pm42$ & $0.10$ & $2.68$ \\
RRabl$16_1$  & $144.36704$ & $+56.59926$ & $0.52854$ & $16.67$ & $0.94$ & $
-5.50E-4\pm9.89E-5$ & $-165\pm42$ & $0.22$ & $2.45$ \\
RRabl$16_2$  & $144.36704$ & $+56.59926$ & $0.52854$ & $16.67$ & $0.94$ & $
-5.11E-4\pm9.89E-5$ & $-153\pm42$ & $0.19$ & $0.76$ \\
RRabl$16_3$  & $144.36704$ & $+56.59926$ & $0.52854$ & $16.67$ & $0.94$ & $
-4.62E-4\pm2.50E-5$ & $-138\pm18$ & $0.28$ & $0.26$ \\
RRabl17  & $144.45373$ & $+0.16152$ & $0.55127$ & $14.67$ & $0.99$ & $
-3.23E-4\pm3.80E-5$ & $-97\pm16$ & $0.18$ & $5.52$ \\
RRabl18  & $145.76492$ & $+10.31705$ & $0.67384$ & $13.15$ & $0.73$ & $
-2.21E-4\pm1.43E-5$ & $-66\pm8$ & $0.25$ & $4.20$ \\
RRabl19  & $147.23862$ & $+17.52836$ & $0.53657$ & $15.85$ & $1.06$ & $
-5.78E-4\pm9.89E-5$ & $-173\pm42$ & $0.16$ & $2.65$ \\
RRabl20  & $148.41207$ & $+26.51218$ & $0.57894$ & $17.06$ & $0.76$ & $
-2.85E-4\pm5.56E-5$ & $-85\pm34$ & $0.30$ & $5.33$ \\
RRabl21  & $149.35586$ & $+3.66790$ & $0.53401$ & $14.41$ & $1.04$ & $
-4.24E-4\pm2.76E-5$ & $-127\pm14$ & $0.09$ & $3.69$ \\
RRabl22  & $156.00410$ & $+23.57355$ & $0.57618$ & $16.23$ & $0.87$ & $
-6.83E-4\pm3.16E-5$ & $-205\pm32$ & $0.21$ & $3.74$ \\
RRabl23  & $156.42426$ & $+27.70770$ & $0.54505$ & $17.72$ & $1.01$ & $
-1.43E-4\pm3.55E-5$ & $-43\pm27$ & $0.43$ & $3.96$ \\
RRabl24  & $158.23468$ & $+9.95462$ & $0.57453$ & $15.13$ & $0.71$ & $
-5.07E-4\pm4.13E-5$ & $-152\pm26$ & $0.13$ & $2.83$ \\
RRabl25  & $160.76055$ & $+6.57957$ & $0.51591$ & $14.31$ & $0.78$ & $
-5.71E-4\pm3.26E-5$ & $-171\pm10$ & $0.07$ & $1.77$ \\
RRabl26  & $163.78099$ & $-3.94871$ & $0.83472$ & $15.20$ & $1.19$ & $
-2.70E-4\pm2.50E-5$ & $-81\pm11$ & $0.13$ & $5.09$ \\
RRabl27  & $172.95127$ & $+16.94800$ & $0.62769$ & $15.56$ & $0.87$ & $
-5.17E-4\pm9.93E-5$ & $-155\pm42$ & $0.11$ & $2.59^{*}$ \\
RRabl28  & $186.22131$ & $+3.89524$ & $0.50837$ & $16.62$ & $0.98$ & $
-6.40E-4\pm9.87E-5$ & $-192\pm42$ & $0.12$ & $1.87$ \\
RRabl29  & $190.79904$ & $+44.92621$ & $0.59282$ & $15.56$ & $0.96$ & $
-2.05E-4\pm5.82E-5$ & $-61\pm34$ & $0.33$ & $10.26$ \\
RRabl30  & $196.40574$ & $+59.99944$ & $0.51326$ & $15.42$ & $1.11$ & $
-4.77E-4\pm3.70E-5$ & $-143\pm14$ & $0.11$ & $3.35$ \\
RRabl31  & $196.74774$ & $+16.74280$ & $0.57767$ & $17.35$ & $0.55$ & $
-5.39E-4\pm5.04E-5$ & $-162\pm19$ & $0.27$ & $2.60$ \\
RRabl32  & $201.57540$ & $+2.64155$ & $0.65792$ & $14.97$ & $0.83$ & $
-3.48E-4\pm3.81E-5$ & $-104\pm29$ & $0.23$ & $3.57$ \\
RRabl33  & $204.94889$ & $+15.59276$ & $0.53250$ & $15.04$ & $1.09$ & $
-4.19E-4\pm4.14E-5$ & $-126\pm19$ & $0.18$ & $4.34$ \\
RRabl34  & $207.17689$ & $+41.91878$ & $0.54509$ & $15.27$ & $0.91$ & $
-5.64E-4\pm9.87E-5$ & $-169\pm42$ & $0.16$ & $2.28$ \\
RRabl35  & $207.84723$ & $+3.96548$ & $0.62922$ & $15.22$ & $1.06$ & $
-5.50E-4\pm8.25E-5$ & $-165\pm25$ & $0.07$ & $3.69$ \\
RRabl36  & $208.18557$ & $-2.30798$ & $0.57549$ & $14.10$ & $1.04$ & $
-4.63E-4\pm2.47E-5$ & $-139\pm10$ & $0.18$ & $2.44$ \\
RRabl37  & $211.07900$ & $-4.10587$ & $0.55437$ & $13.83$ & $0.90$ & $
-6.13E-4\pm9.89E-5$ & $-184\pm42$ & $0.05$ & $2.36$ \\
RRabl38  & $219.15778$ & $+14.79704$ & $0.52376$ & $15.85$ & $1.00$ & $
-6.19E-4\pm2.74E-5$ & $-186\pm10$ & $0.18$ & $1.02$ \\
RRabl39  & $221.29128$ & $+1.50971$ & $0.67651$ & $14.97$ & $0.89$ & $
-6.69E-4\pm9.93E-5$ & $-200\pm42$ & $0.08$ & $2.42$ \\
RRabl40  & $221.37085$ & $-4.02095$ & $0.51691$ & $13.68$ & $0.96$ & $
-6.04E-4\pm2.43E-5$ & $-181\pm11$ & $0.12$ & $3.90$ \\
\hline
\end{tabular}}
%\end{longtable}}
%\end{supertabular}}
\end{center}
\end{table*}

\begin{table*}[tbp]
%\centering
\begin{center}
\setlength{\tabcolsep}{1mm}{
\begin{tabular}{lccccccccccccccc}
\hline
{Object}   & R.A.(J2000) & Decl.(J2000) & Period & $V$ & Amp & $z_{\rm
e1,\alpha}$ & $V_{\rm e1,\alpha}$ &  Flux$_{\rm e1,\alpha}$ & FWHM$_{\rm
e1,\alpha}$\\
$  $          & ${}^{\circ}$ & ${}^{\circ}$ & day & mag & mag        & $ $ &
km/s &   & $\rm \mathring{A}$   \\
\hline
RRabl41  & $224.32736$ & $+22.98642$ & $0.57342$ & $15.42$ & $0.92$ & $
-6.17E-4\pm9.91E-5$ & $-185\pm42$ & $0.07$ & $2.56$ \\
RRabl42  & $233.40562$ & $+0.43740$ & $0.51954$ & $15.78$ & $0.82$ & $
-4.45E-4\pm9.95E-5$ & $-133\pm42$ & $0.11$ & $2.86$ \\
RRabl43  & $235.23457$ & $+10.96986$ & $0.58815$ & $15.29$ & $0.79$ & $
-6.42E-4\pm4.46E-5$ & $-192\pm14$ & $0.06$ & $2.28$ \\
RRabl44  & $239.16098$ & $+4.04318$ & $0.55646$ & $14.49$ & $0.81$ & $
-5.68E-4\pm9.89E-5$ & $-170\pm42$ & $0.11$ & $2.52$ \\
RRabl45  & $239.38257$ & $+28.63362$ & $0.66540$ & $12.51$ & $0.53$ & $
-7.04E-4\pm1.54E-5$ & $-211\pm9$ & $0.14$ & $3.18$ \\
RRabl46  & $240.30956$ & $+2.61588$ & $0.66357$ & $14.72$ & $1.00$ & $
-6.70E-4\pm3.41E-5$ & $-201\pm22$ & $0.16$ & $1.61$ \\
RRabl47  & $240.35335$ & $+27.02554$ & $0.52700$ & $14.17$ & $0.84$ & $
-7.82E-4\pm2.51E-5$ & $-235\pm11$ & $0.14$ & $3.78$ \\
RRabl48  & $242.48518$ & $+6.42657$ & $0.64348$ & $15.75$ & $0.88$ & $
-1.74E-4\pm2.08E-5$ & $-52\pm15$ & $0.25$ & $4.26$ \\
RRabl49  & $242.93347$ & $+17.00636$ & $0.49055$ & $14.76$ & $1.04$ & $
-4.70E-4\pm7.38E-5$ & $-141\pm29$ & $0.19$ & $1.31$ \\
RRabl50  & $245.90139$ & $+36.42249$ & $0.57302$ & $14.38$ & $1.12$ & $
-4.80E-4\pm9.91E-5$ & $-144\pm42$ & $0.08$ & $1.70$ \\
RRabl51  & $249.52973$ & $-3.43423$ & $0.56249$ & $14.86$ & $1.00$ & $
-4.67E-4\pm9.91E-5$ & $-140\pm42$ & $0.10$ & $0.98$ \\
RRabl52  & $250.09389$ & $+15.06886$ & $0.69196$ & $13.95$ & $0.74$ & $
-5.01E-4\pm3.18E-5$ & $-150\pm10$ & $0.15$ & $0.75$ \\
RRabl53  & $253.42009$ & $+34.38424$ & $0.61463$ & $15.56$ & $0.53$ & $
-5.78E-4\pm9.93E-5$ & $-173\pm42$ & $0.16$ & $1.34$ \\
RRabl54  & $254.49119$ & $+30.71234$ & $0.54937$ & $16.84$ & $0.91$ & $
-1.69E-4\pm6.38E-5$ & $-51\pm33$ & $0.43$ & $3.93$ \\
RRabl55  & $259.42280$ & $+41.32834$ & $0.57087$ & $14.77$ & $0.86$ & $
-6.95E-4\pm1.82E-5$ & $-208\pm8$ & $0.15$ & $3.43$ \\
RRabl56  & $261.66390$ & $+40.15591$ & $0.63667$ & $17.75$ & $0.94$ & $
-1.03E-4\pm5.16E-5$ & $-31\pm31$ & $0.41$ & $7.09$ \\
RRabl57  & $266.22514$ & $+42.81815$ & $0.51673$ & $15.43$ & $1.02$ & $
-2.46E-4\pm3.86E-5$ & $-74\pm15$ & $0.17$ & $4.38$ \\
RRabl58  & $319.48228$ & $-3.03425$ & $0.61376$ & $15.60$ & $0.44$ & $
-9.01E-4\pm9.91E-5$ & $-270\pm42$ & $0.13$ & $0.91$ \\
RRabl59  & $347.44815$ & $+29.67568$ & $0.60791$ & $16.68$ & $0.63$ & $
-5.50E-4\pm9.89E-5$ & $-165\pm42$ & $0.07$ & $1.72$ \\
RRabl60  & $352.58414$ & $+21.11239$ & $0.52758$ & $14.91$ & $0.97$ & $
-5.88E-4\pm1.74E-5$ & $-176\pm7$ & $0.17$ & $2.44$ \\
RRabl61  & $357.69274$ & $+33.35102$ & $0.53392$ & $13.20$ & $0.77$ & $
-6.29E-4\pm2.33E-5$ & $-189\pm15$ & $0.19$ & $3.95$ \\
\hline
{Object}   & R.A.(J2000) & Decl.(J2000) & Periodf & $G$ & AmpG & $z_{\rm
e1,\alpha}$ & $V_{\rm e1,\alpha}$ &  Flux$_{\rm e1,\alpha}$ & FWHM$_{\rm
e1,\alpha}$\\
$  $          & ${}^{\circ}$ & ${}^{\circ}$ & day & mag & mag        & $ $ &
km/s &   & $\rm \mathring{A}$   \\
\hline
RRabl62  & $10.49515$ & $+33.83455$ & $0.50996$ & $16.89$ & $0.91$ & $
-6.51E-4\pm1.49E-5$ & $-195\pm10$ & $0.31$ & $2.72$ \\
RRabl$63_1$  & $178.36094$ & $+53.77534$ & $0.54987$ & $17.38$ & $0.81$ & $
-5.60E-4\pm2.04E-5$ & $-168\pm11$ & $0.38$ & $2.03$ \\
RRabl$63_2$  & $178.36094$ & $+53.77534$ & $0.54987$ & $17.38$ & $0.81$ & $
-5.72E-4\pm9.89E-5$ & $-171\pm42$ & $0.26$ & $2.49$ \\
RRabl64  & $180.93050$ & $+5.48829$ & $0.57846$ & $15.58$ & $1.05$ & $
-3.50E-4\pm3.28E-5$ & $-105\pm15$ & $0.19$ & $2.51$ \\
RRabl65  & $196.90430$ & $+73.59784$ & $0.51574$ & $15.04$ & $0.97$ & $
-4.89E-4\pm2.84E-5$ & $-147\pm12$ & $0.14$ & $2.89$ \\
RRabl66  & $248.94414$ & $-4.75558$ & $0.66401$ & $16.06$ & $0.67$ & $
-1.35E-4\pm9.81E-5$ & $-41\pm42$ & $0.59$ & $3.47^{*}$ \\
RRabl67  & $285.76599$ & $+57.32673$ & $0.52887$ & $14.43$ & $0.88$ & $
-2.74E-4\pm2.58E-5$ & $-82\pm13$ & $0.11$ & $5.66$ \\
RRabl68  & $291.08516$ & $+34.27314$ & $0.56338$ & $15.12$ & $0.72$ & $
-5.80E-4\pm5.86E-5$ & $-174\pm18$ & $0.06$ & $2.12$ \\
RRabl69  & $324.40104$ & $+44.46210$ & $0.47481$ & $16.30$ & $0.87$ & $
-5.72E-4\pm9.91E-5$ & $-171\pm42$ & $0.14$ & $1.48$ \\
RRabl$70_1$  & $326.54851$ & $+26.93884$ & $0.57201$ & $13.73$ & $0.89$ & $
-6.67E-4\pm2.53E-5$ & $-200\pm10$ & $0.14$ & $3.17$ \\
RRabl$70_2$  & $326.54851$ & $+26.93884$ & $0.57201$ & $13.73$ & $0.89$ & $
-2.49E-4\pm1.43E-5$ & $-75\pm9$ & $0.17$ & $4.63$ \\

\hline
{Object}   & R.A.(J2000) & Decl.(J2000) & Period & $V$ & Amp & $z_{\rm
e1,\alpha}$ & $V_{\rm e1,\alpha}$ &  Flux$_{\rm e1,\alpha}$ & FWHM$_{\rm
e1,\alpha}$\\
$  $          & ${}^{\circ}$ & ${}^{\circ}$ & day & mag & mag        & $ $ &
km/s &   & $\rm \mathring{A}$   \\
\hline
RR-Bl  & $322.09646$ & $+20.22167$ & $0.54283$ & $14.84$ & $0.41$ & $
-2.55E-4\pm9.83E-5$ & $-76\pm42$ & $0.12$ & $3.98$ \\
\hline
\end{tabular}}
%\end{longtable}}
%\end{supertabular}}
\end{center}
\tablecomments{\\\hspace{\textwidth}
1. Period, $V$ and Amp (Amplitude) are produced by Catalina Sky Survey.
Periodf, $G$ and AmpG are produced by {Gaia} mission. Periodf
denotes fundamental period in the {Gaia} catalog.\\\hspace{\textwidth}
2. $z_{\rm e1,\alpha}$ represents the redshift of the emission component of the
``first apparition'' in the stellar rest frame.\\\hspace{\textwidth}
3. $V_{\rm e1,\alpha}$ represents the radial velocity of the emission component
of the ``first apparition'' in the stellar rest frame.\\\hspace{\textwidth}
4. Flux$_{\rm e1,\alpha}$ indicates the normalized flux of the emission.
\\\hspace{\textwidth}
5. FWHM$_{\rm e1,\alpha}$ indicates full width at half maximum of the emission.
\\\hspace{\textwidth}
6. The names like RRabl$5_1$ mean that there are not only one spectrum of one
star. \\\hspace{\textwidth}
7. The measurements with $^*$ denote low-quality results.\\\hspace{\textwidth}
8. The stars are ordered by R.A.(J2000).}
 \end{table*}

\begin{figure*}[tb]
%\centering
\begin{center}
\begin{tabular}{c}
%\centering
\includegraphics*[width=120mm,height=8cm]{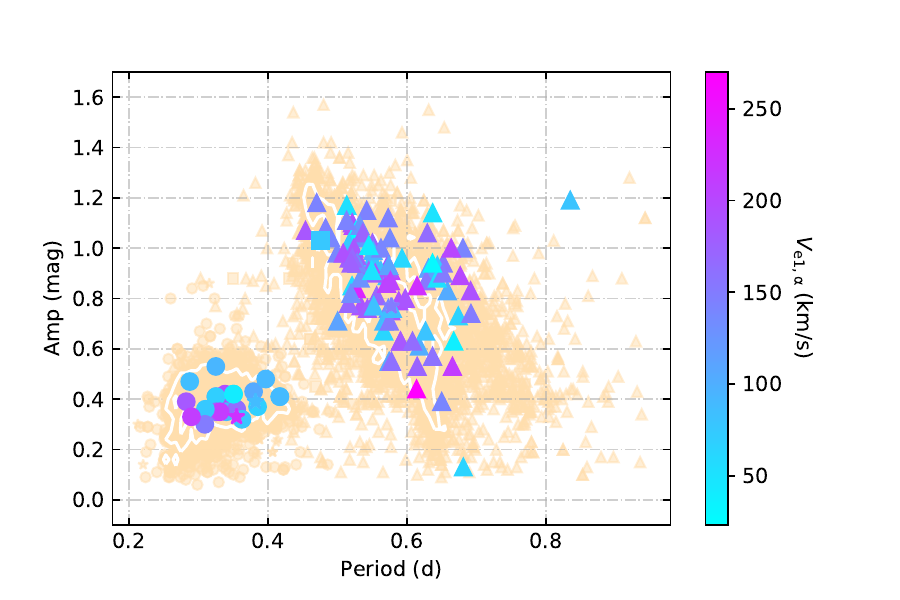}
\end{tabular}
\caption{\label{fig:periodamp} Period-Amplitude diagram. Selected stars with
period and amplitude produced by Catalina Sky Survey are shown. Corresponding
relation: triangle - RRab, circle - RRc, star - RRd, square - the Blazhko type
RR Lyrae variable. The variation of color shows different absolute value of
$V_{\rm e1,\alpha}$. Light yellow points in the background denote the whole
sample of RR Lyrae stars. $1\sigma$ regions are shown as white contours.}
%\end{tabular}
\end{center}
\end{figure*}

\begin{figure*}[tb]
\begin{center}
\begin{tabular}{c}
\includegraphics*[width=180mm,height=16.2cm]{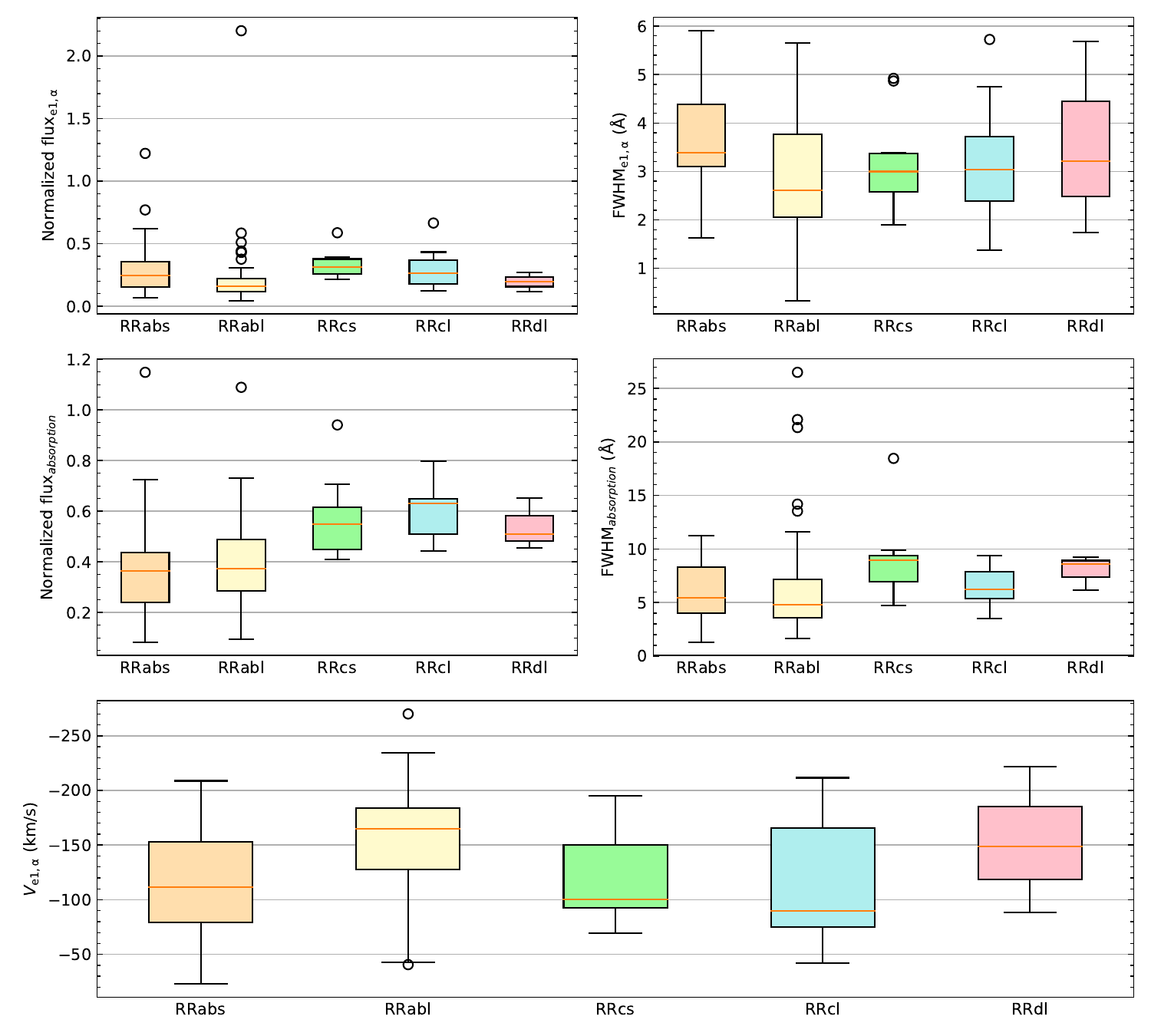}
\end{tabular}
\caption{\label{fig:distributionofparameter}
Box plots of measurements of different type of stars that show the ``first
apparitions''. Upper left (a): Normalized flux of emission (Flux$_{\rm
e1,\alpha}$); Upper right (b): FWHM of emission (FWHM$_{\rm e1,\alpha}$); Lower
left (c): Normalized flux of absorption; Lower right (d): FWHM of absorption.
Bottom (e): $V_{\rm e1,\alpha}$ of different type of stars that show the
``first apparition''.}\end{center}
\end{figure*}

The distribution of selected stars on Period-Amplitude diagram is shown in
Figure~\ref{fig:periodamp}. A clear gap divides RRab and RRc stars into two
groups. RRab stars have longer periods and larger mean amplitudes than RRc
stars.
Box plots of the measurements are displayed in
Figure~\ref{fig:distributionofparameter}. The mean value of normalized flux
or FWHM of emission or absorption of RRc stars are higher than those of RRab
stars. The mean value of $V_{\rm e1,\alpha}$ of RRab stars is higher than RRc
stars.

\begin{figure*}[tb]
\begin{center}
\begin{tabular}{c}
\includegraphics*[width=90mm,height=6cm]{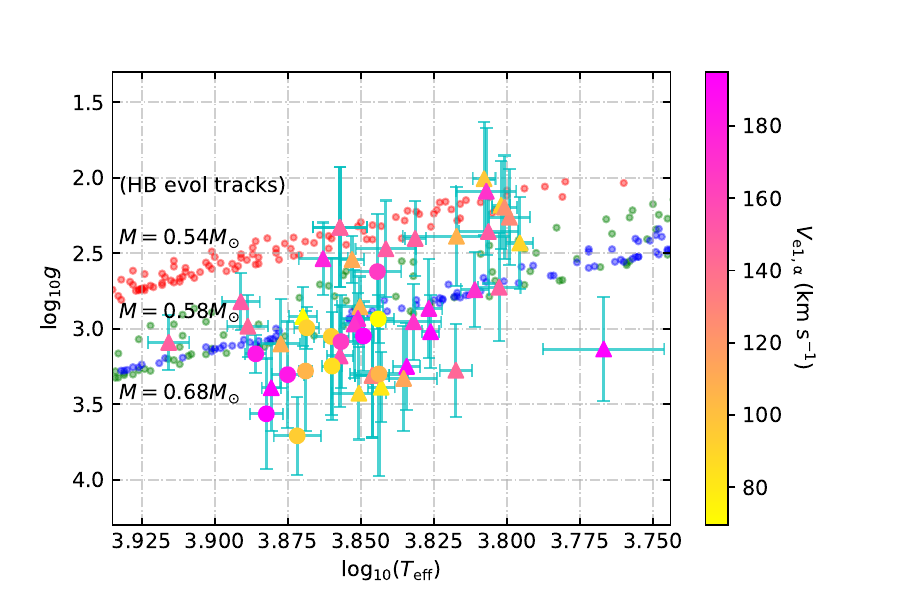}
\end{tabular}
\caption{\label{fig:logglogTobservation} log $T_{\rm eff}$-log\,$g$ diagram for
RR Lyrae stars that show the ``first apparition''. Measurements of $T_{\rm
eff}$ and log\,$g$ were produced by SSPP, if available. Points with error bars
denote the selected stars with the ``first apparitions''. Triangles denote RRab
stars while circles denote RRc stars.
The variations of color show different $V_{\rm e1,\alpha}$. The line consisting
of red, green, and blue points in the background indicate the simulated
horizontal-branch (HB) evolutionary tracks for stars with $M = 0.54M_{\odot},
0.58M_{\odot}, 0.68M_{\odot}$, respectively \citep{Dorman1992ApJS...81..221D}.}
\end{center}
\end{figure*}

Figure~\ref{fig:logglogTobservation} shows the log $T_{\rm eff}$-log\,$g$
diagram.  We adopt measurements of $T_{\rm eff}$ and log\,$g$ generated by SSPP if available. Despite the pipeline is not optimized for RR Lyrae stars
and these values are derived from co-added spectra, the parameters from the
pipeline can provide an overall description of log\,$g$-log $T_{\rm eff}$ distribution of
various subgroups. Overall, RRc stars are hotter and have larger log\,$g$ than
RRab stars. We also display simulated horizontal-branch (HB) evolutionary
tracks for stars in the background, which was calculated with expansion and
semiconvection of the core and enhanced oxygen composition from
\cite{Dorman1992ApJS...81..221D}, to show regions of different masses. The line
consisting of red, green, and blue points in the background indicates
evolutionary tracks for stars with $M = 0.54, 0.58, 0.68M_{\odot}$,
respectively.

\begin{figure*}[tb]

\begin{center}
\begin{tabular}{cc}

\includegraphics*[width=90.0mm,height=6cm]{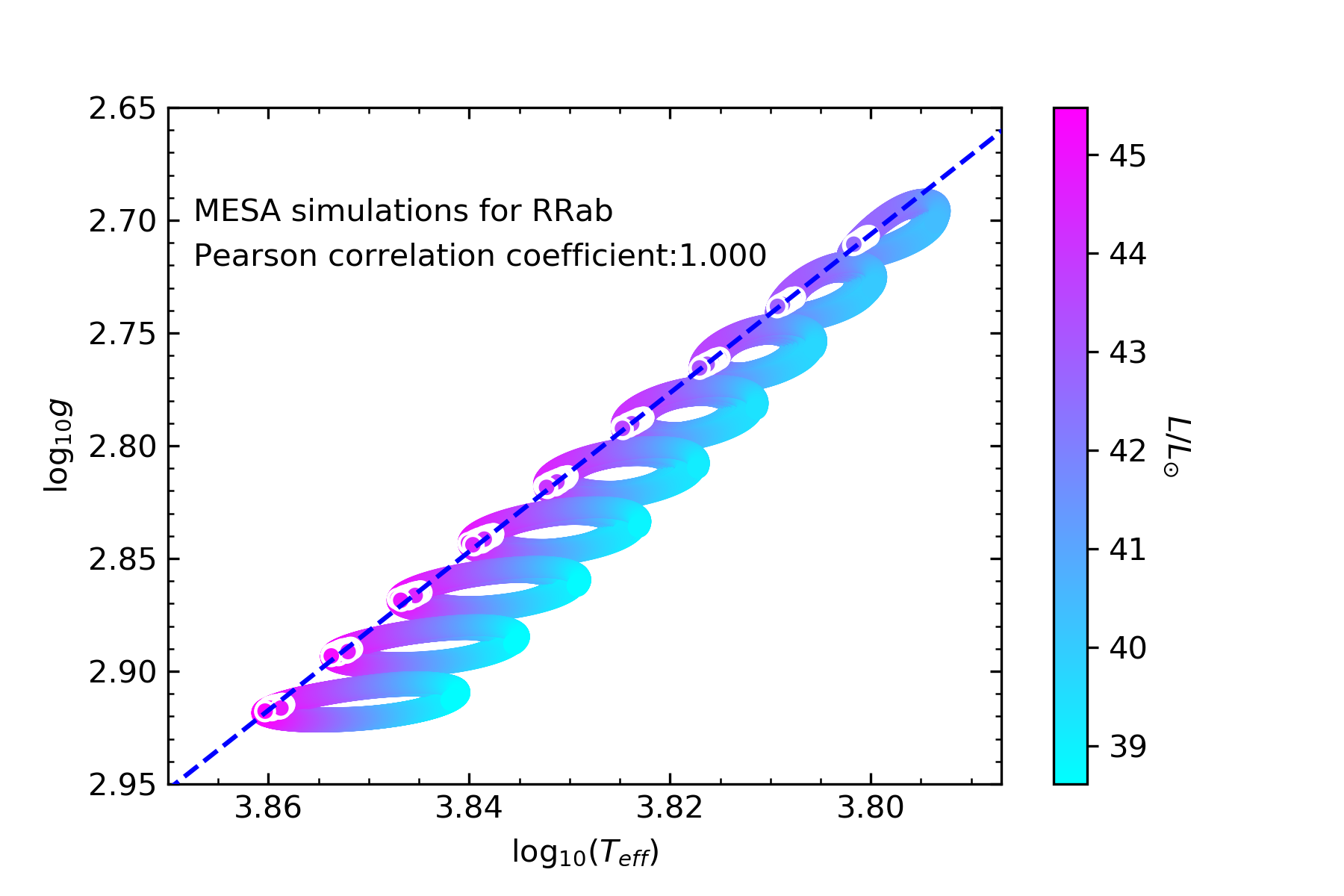}&
\includegraphics*[width=90.0mm,height=6cm]{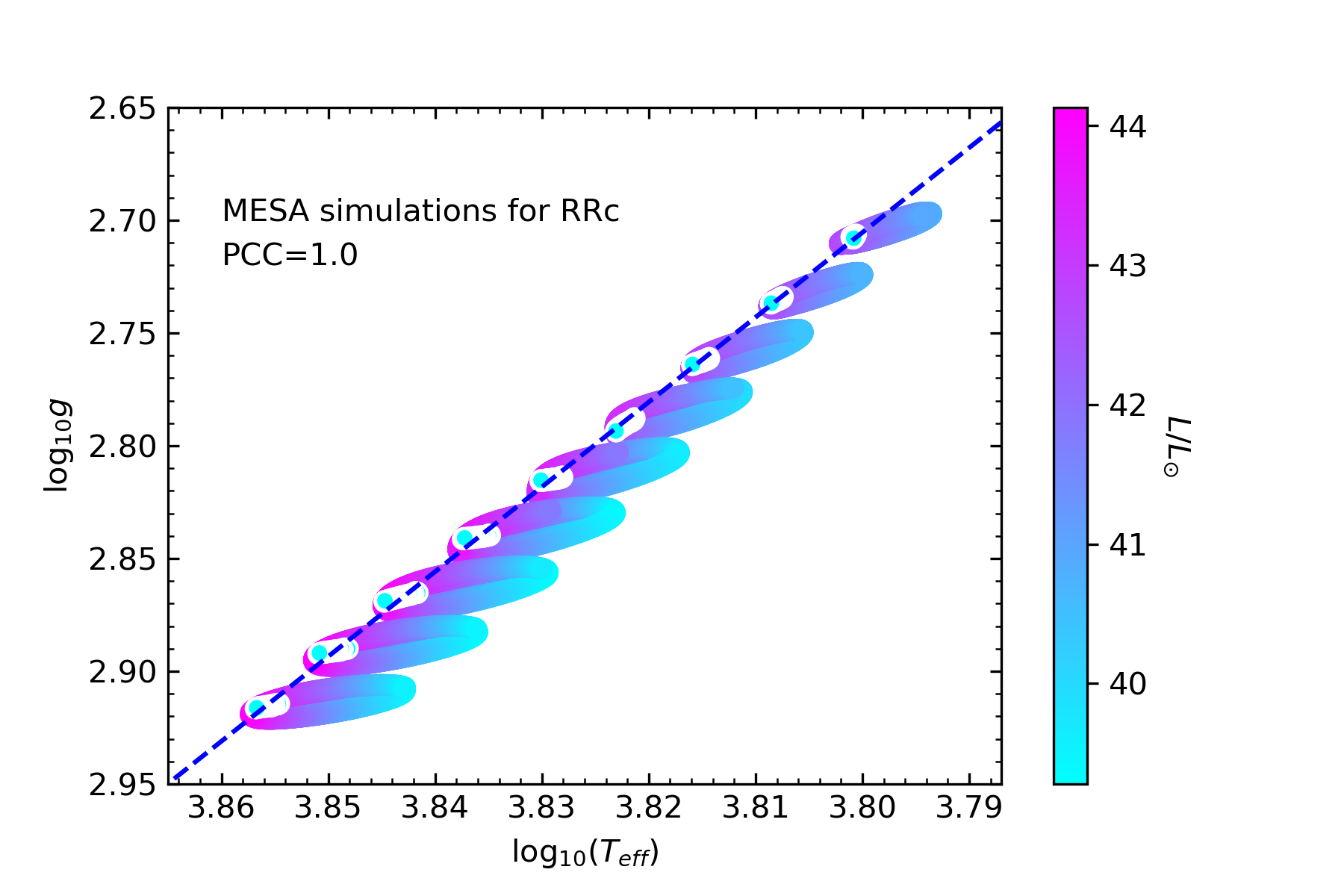}
\end{tabular}
\caption{\label{fig:MESAsimulation} Left (a): log $T_{\rm eff}$-log\,$g$ diagram
generated from MESA simulation for RRab and RRc stars. Left (a): Evolutionary
tracks for a series of RRab stars during identical time intervals. The
variations of color denote different $L/L_{\odot}$. The white points indicate
the locations of $\phi=0.91$. The black dashed line show the linear fitting for
the white points. As for initial parameters, $M_{init}=0.55M_{\odot}$,
$L_{init}=41.687L_{\odot}$, $X_{init}=0.736$ and $Z_{init}=0.008$, which
indicates $[Fe/H]=-0.33$. Initial $T_{\rm eff}$ ranges from 6300K to 7100K in
step width of 100K. 
Right (b): Evolutionary tracks for a series of RRc stars during identical time
intervals. Other comments are the same as Figure~9a.}
\end{center}
\end{figure*}

The presence of the ``first apparition'' indicates a certain range of
phase, just before the maximum luminosity.
Figure~\ref{fig:MESAsimulation} shows that points during a certain phase
interval from RR Lyrae variables with similar properties is concentrating on a
linear path on log $T_{\rm eff}$-log\,$g$ diagram when the effective temperature
changes, which is demonstrated by a series of simulation using Modules for
Experiments in Stellar Astrophysics \citep[MESA,][]{Paxton2019ApJS..243...10P}.
We simulated 9 RRab and 9 RRc stars with various initial effective temperature
in identical duration of evolution, while other initial parameters remain the
same. The initial parameters are $M_{init}=0.55M_{\odot}$,
$L_{init}=41.687L_{\odot}$, $X_{init}=0.736$ and $Z_{init}=0.008$, which
indicates $[Fe/H]=-0.33$ \citep{Das2018MNRAS.481.2000D}. $T_{eff}$ ranges from 6300K to 7100K in step width of 100K. In Figure~\ref{fig:MESAsimulation}, variation of color
indicates the changing of luminosity, which can stand for the changing phase.
Points at around $\phi=0.91$ locate on a
linear path, highly characterized by Pearson correlation coefficients (PCC) as
1.000 and 0.999 for RRab and RRc respectively. The observation points on log
$T_{\rm eff}$-log\,$g$ diagram have a diffuse distribution because our selected
stars have various physical properties such as mass, luminosity, and
metallicity. The different parameters make them deviate from one linear path.

\section{Analysis of frequency components in the RR-Bl}\label{section:Blazhko}

\begin{figure*}[tb]
\begin{center}
\begin{tabular}{c}
\includegraphics*[width=180mm,height=16.2cm]{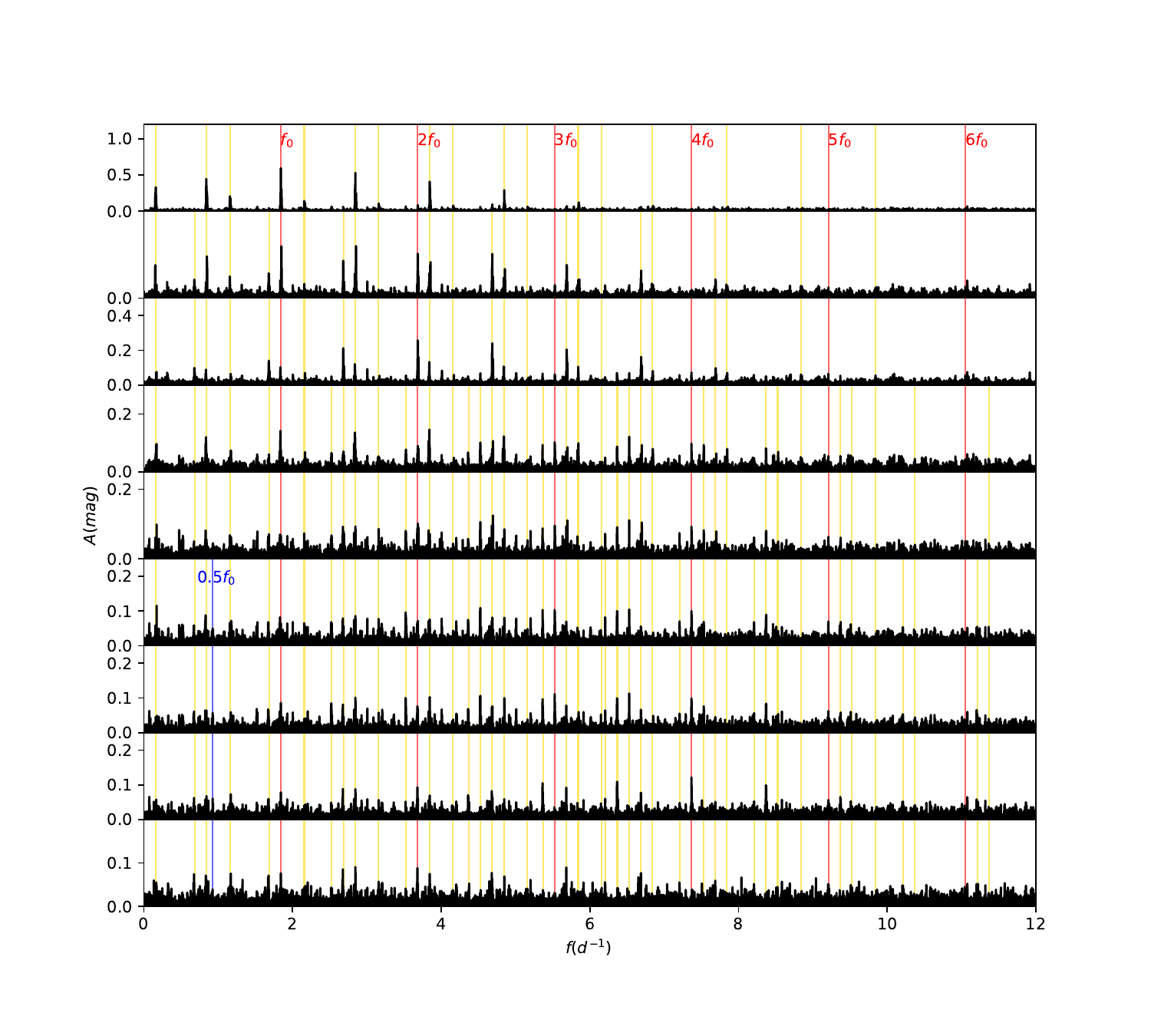}
\end{tabular}
\caption{\label{fig:prewhitening} Pre-whitening sequence for RR-Bl. Uppermost
panel displays power spectrum of the original data. Lower eight panels show
power spectra after removing consecutive frequencies. $nf_0$ lines are marked
with red color. $0.5f_0$ is marked with blue color. Yellow lines are linear
combinations expressed by $nf_0+\lambda_cf_c$, while $f_c$ is the daily
cadence.}
\end{center}
\end{figure*}

\begin{figure*}[tb]
\begin{center}
\begin{tabular}{c}
\includegraphics*[width=120mm,height=8cm]{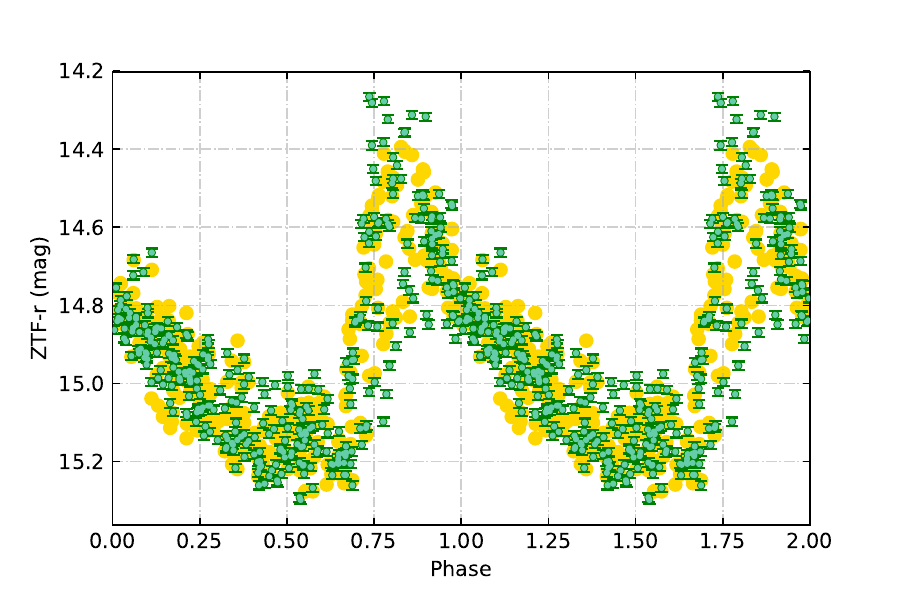}
\end{tabular}
\caption{\label{fig:newlightcurve} Regeneration of light curve of RR-Bl
with full light curve solution. Green points with error bars are photometric
observations from ZTF-r. Golden points are regenerated points with full light
curve solution.}\end{center}
\end{figure*}

We analyze $r$-band light curve of RR-Bl with a time span of 974.66 days with a
standard successive pre-whitening method \citep{Moskalik2009MNRAS.394.1649M}.
Observations in poor quality (catflags $\neq$ 0) have been removed. We fit the
data using a non-linear least-square procedure with the sine series of the
following form at each step:
\begin{eqnarray}\label{eq:sineseries}
m(t) = m_0 + \sum^N_{k=1} A_k {\rm sin}(2\pi f_k t+\phi_k),
\end{eqnarray}
where $f_k$ denotes independent frequencies detected in discrete Fourier
transform of the data and their possible linear combinations. The pre-whitening
sequence for RR-Bl is shown in Figure~\ref{fig:prewhitening}. We find a $full$
$light$ $curve$ $solution$ for RR-Bl and produce the fitting results of
the light curve in Figure~\ref{fig:newlightcurve}.

According to $f_0 = 1.84226$, we derive the fundamental period $P_0 = 0.54281
d$ and $A_0 = 0.242$ mag. Apart from $f_0$ and its linear combinations $nf_0$,
we detect significant signals that can be expressed as $nf_0+\lambda_cf_c$,
where $f_c\approx1$, which can be explained as daily cadence. According to the
pre-whitening sequence for RR-Bl, only one independent dominant frequency is
detected. So RR-Bl is an RRab star with strong Blazhko effect.

\begin{figure*}[tb]
\begin{center}
\begin{tabular}{c}
\includegraphics*[width=180mm,height=10cm]{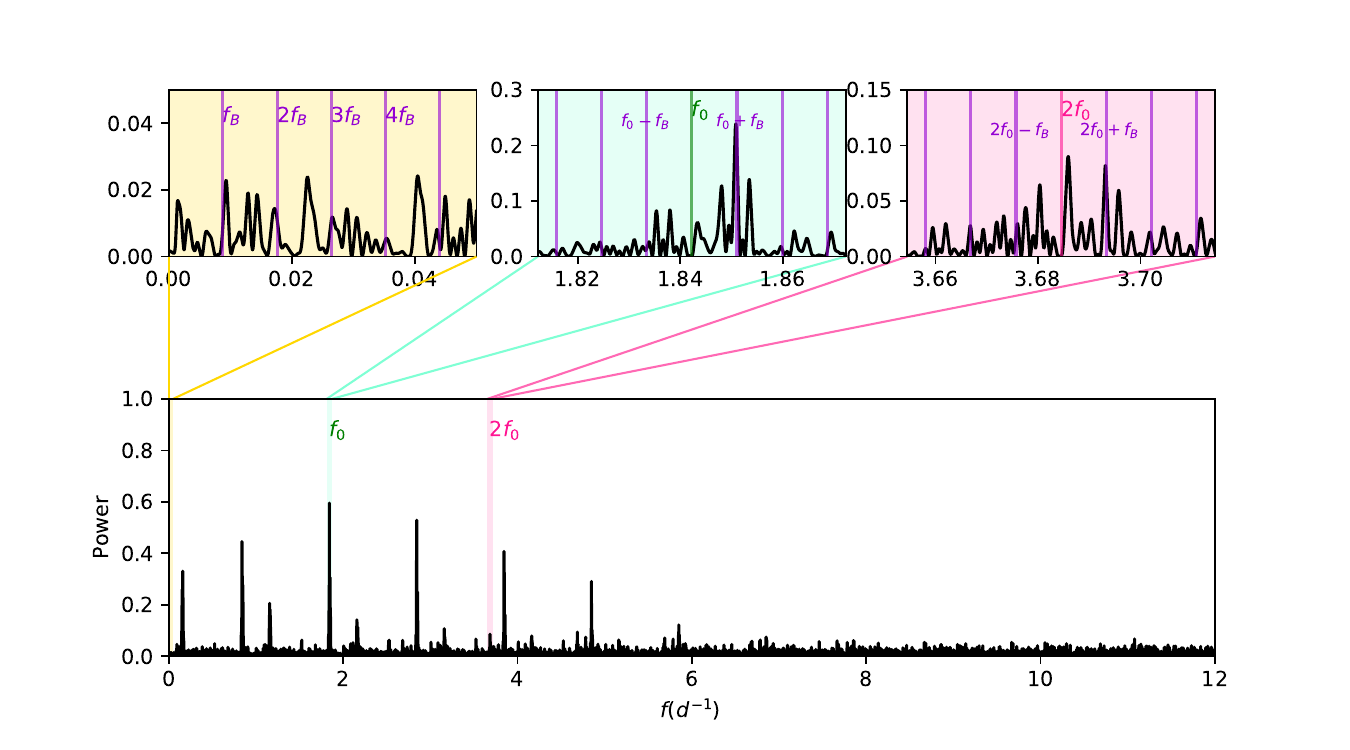}
\end{tabular}
\caption{\label{fig:Blazoom} Pre-whitened power spectrum of RR-Bl.
Lower panel displays power spectrum of the original data. Upper panels display
power spectra after pre-whitening, showing the fine structures in low
frequency, around $f_0$, and $2f_0$, respectively.}
\end{center}
\end{figure*}

The additional close side peaks at the fundamental frequency denote long-term
modulation \citep{Smolec2015MNRAS.447.3756S}. In Figure~\ref{fig:Blazoom}, the
amplifying pattern is shown. The weak signatures beside $nf_0$ indicate a
possible Blazhko-type modulation, with $f_B\approx0.009$ and
$P_B\approx111.111{\rm d}$. The signals other than $f_0+f_B$ are not enough
significant compared to others. Larger datasets are needed for a more precise
analysis.

\section{Detection rate and mock data test}\label{section:detectionrateproblem}

The observational detection rate of blueshifted emission features in hydrogen lines of RR Lyrae stars is defined as: 
\begin{eqnarray}\label{eq:detection rate3}
\eta = \frac{\rm{N}_{obs}}{\rm{N}_{sat}},
\end{eqnarray}
where $\rm N_{\rm obs}$ represents the number of RRab stars featured emission components in LAMOST survey spectra. $\rm{N}_{sat}$ denotes the input number of RRab variables in the survey plan that are observed and with the S/N ratios of spectra greater than 15. Such definition of detection rate has an implicit assumption, that all RRab stars possess shock wave triggered emission, may or may not be detected in random sampling spectroscopic observations, depending on the observational timing and the strength of the emission feature.

The results of $\eta$ are: 1.84$\%$ - RRab in SDSS, 1.06$\%$ - RRc in SDSS, 2.87$\%$ -
RRab in LAMOST, 1.45$\%$ - RRc in LAMOST, 2.97$\%$ - RRd and 2.56$\%$ for
Blazhko type RR Lyrae star in LAMOST. The detection rate of the ``first
apparition'' of RRc stars are significantly lower than RRab stars. We suggest that the detection rate in LAMOST are higher than those in SDSS is because that the sources in SDSS are fainter and the pixel resolution of LAMOST is higher than SDSS. As for SDSS, we choose the ``spCFrame spectra'', the pixel resolutions are about 4150 at H$\alpha$ and 4180 at H$\beta$. As for LAMOST, the pixel resolutions are about 7800 at H$\alpha$ and 8400 at H$\beta$. The pixel resolution influences our results because one selecting criterion is that both emission and absorption components should contain at least two observational points. We set up two simulations for RRab variables in LAMOST: one on mock spectra to test the performance of our pipeline; the other on the observational strategy to investigate the occurrence of the ``first apparition''.

Firstly, we test the performance of our pipeline on mock spectra generated based on the distributions of real observational parameters. We propose that the recognition rate of the shock wave signatures can be described as:
\begin{eqnarray}\label{eq:detection rate2}
P(A)P(B\mid A)=P(AB),
\end{eqnarray}
where A = ``the first apparition is observed on the spectra'', B = ``the signal meets the
criteria''. We recognize $P(A)$ as the real detection rate, while $P(AB)$ is the
detection rate from the search. $P(AB)$ is lower than $P(A)$ due to finite S/N and our strict criteria of selecting the stars with the ``first apparition'', which is described in Section~\ref{section:methods}.

\begin{figure*}[tb]
\begin{center}
\begin{tabular}{c}
\includegraphics*[width=180mm,height=10.8cm]{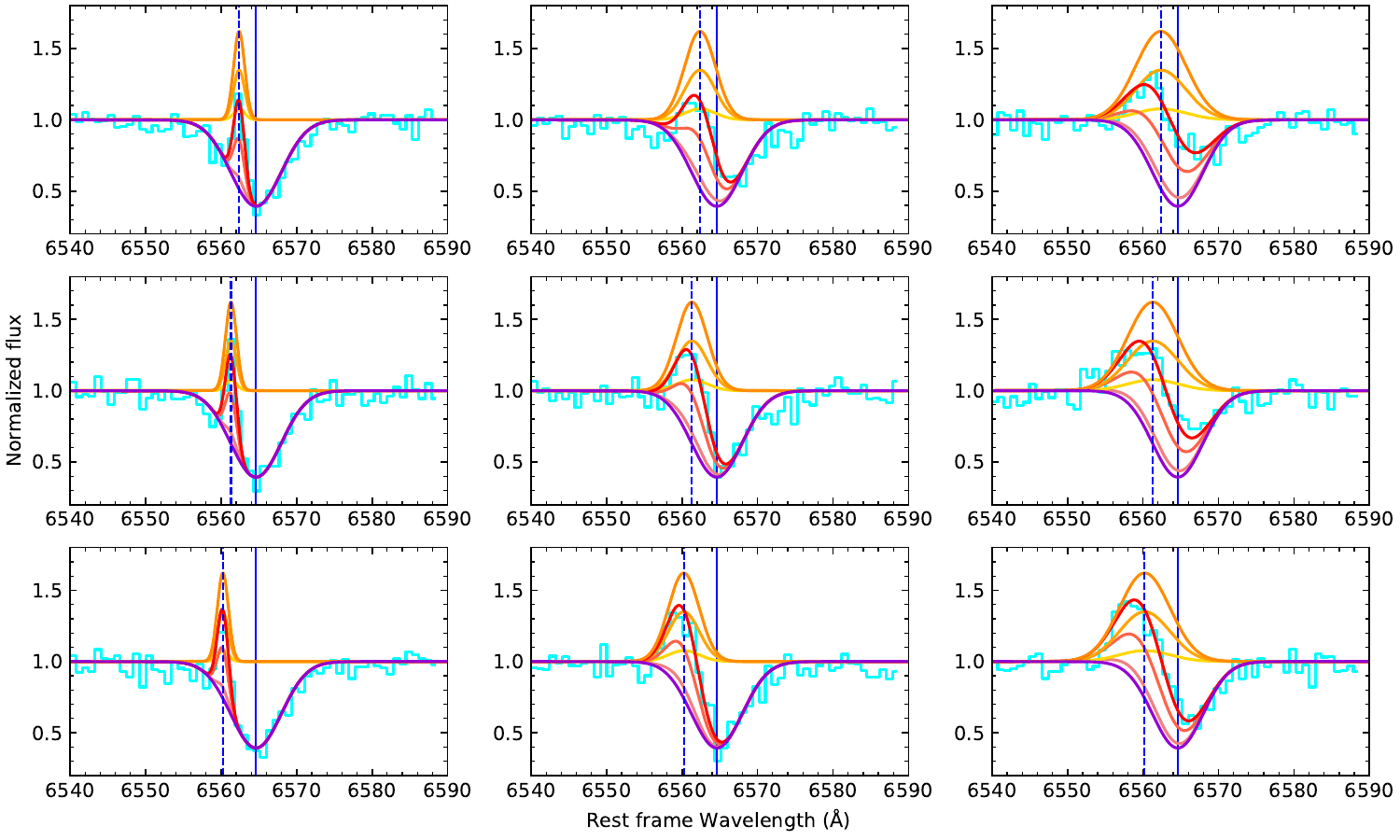}
\end{tabular}
\caption{\label{fig:mockspectra}
Simulated template of spectra of the ``first apparitions''. The wavelength axis
is in the stellar rest frame. The blueshifted H$\alpha$ emission lines are
shown as orange profiles. The H$\alpha$ absorption lines are shown as blue
profiles. Red profiles show the shapes of the ``first apparitions''. Cyan
profiles display the profile with random noises. Vertical blue solid lines
denote the H$\alpha$ line laboratory wavelength. Vertical blue dashed lines
denote the central wavelength of the emissions.}\end{center}
\end{figure*}

We generate the mock spectra that have the ``first apparition'' and
apply our 1D pattern recognition pipeline on the mock sample to estimate
$P(B\mid A)$. The parametrizations of the simulated ``first apparitions'' are shown in
Figure~\ref{fig:mockspectra}. The mock spectrum is regenerated as two
Gaussian-like profiles between 6,540-6,590 $\rm \mathring{A}$. The central line
of the absorption component is fixed at the rest wavelength of H$\alpha$. The
blueshift of the emission, the ratio between the flux of emission and
absorption component and the ratio between FWHM of emission and absorption
component determine the shape of the simulated spectrum. The parameters of the
Gaussian profiles are set as random values under Gaussian distribution based on
the fitted distributions of the parameters of the selected RRab stars in
LAMOST. Random noises are added up to the profiles to reproduce real
situations, under Gaussian distribution where $\sigma$ is the error of
normalized flux. The resolutions and S/Ns of mock spectra are taken from real
observations.

\begin{figure*}[tb]
\begin{center}
\begin{tabular}{c}
\includegraphics*[width=120mm,height=8cm]{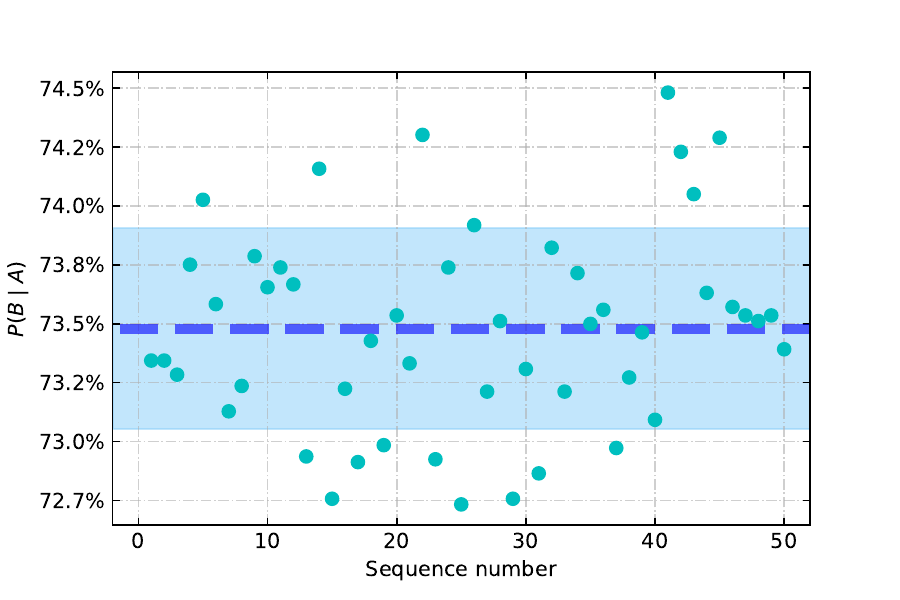}
\end{tabular}
\caption{\label{fig:detectionrate} $P(B\mid A)$ of RRab stars in LAMOST.}
\end{center}
\end{figure*}

We apply our pipeline on the mock sample and set up this simulation for
50 times. We only adopt the result of RRab stars in LAMOST. Because the pixel resolution of SDSS is low and the search in SDSS relies more on eye-checking.
Moreover, the size of the sample of RRc, RRd and Blazhko type RR Lyrae star is
not large enough to provide reliable parameter distribution.
Figure~\ref{fig:detectionrate} displays the results that $P(B\mid A)$ =
73.48\% $\pm$ 0.43\%. This means that, for a spectrum exhibiting blueshifted hydrogen emission with random flux, FWHM, and blueshift, there should be 73.48\% $\pm$ 0.43\% chance that our algorithm would recognize.

Secondly, we set up a following simulation on the real observational strategy to investigate the occurrence of ``first apparition''. The simulation is based on the real parameters, including detection mode, exposure time, observation interval, observing frequency and the period of the corresponding RR Lyrae star for each spectra satisfying S/N > 15.

In the simulation we preset that every star in the testing sample would have a ``first apparition''. The exposure time for a spectrum with S/N > 15 is then used as an observing window. Exposure times and observation intervals are obtained from real spectra. In the simulation, we set that the ``first apparition'' shows up for a short period of time before maximum luminosity in each pulsation cycle. \cite{Chadid2011Carnegie} reported that the ``first apparition'' appears over $\sim 5\%$ of the whole pulsation cycle. \cite{Gillet2019} reported that the ``first apparition'' presents during $\phi=0.892-0.929$ in RR Lyr, accounting for about $3.7\%$ of the whole period. Their researches are based on bright RR Lyrae stars. Our sample is much larger and contains a lot of faint stars. In the simulation, we assume that the ``first apparition'' accounts for about $4\%$ of the whole period. 

As a matter of fact, too long-time exposures can smooth out the emission features in the hydrogen lines. Therefore we assume that the survival time of the emission feature and the exposure time should overlap with at least 70$\%$ of the exposure time for a clear signal. Each case that the emission overlaps with more than 70$\%$ of the exposure time yields one valid catch for a star. The number of valid catches of one star changes with observation time. The possibility of detecting a star with the ``first apparition'' is $P(B\mid A)$, $[1-(1-P(B\mid A))^2]$, or $[1-(1-P(B\mid A))^3]$ during the time when the star has one, two, or three valid catches, respectively.

 As for the result, the number of detected RRab stars with blueshifted hydrogen emission in LAMOST from the simulation is about 371.79, while our observational result is 70. The result of the simulation is larger than our observational result. In our simulation we assume that every star in our sample exhibits the ``first apparition''. However, not every RRab star has a visible ``first apparition'' in real survey due to variant shock strength. The occurrence of ``first apparition'' can be estimated as:
\begin{eqnarray}\label{eq:detection rate1}
P(O)\rm{N}_{theo} = \rm{N}_{obs},
\end{eqnarray}
where O = ``the star shows the first apparition''. We define $P(O)$ as the occurrence of ``first apparition''. $\rm N_{\rm theo}$ is the number of detected RRab stars with blueshifted hydrogen emission in LAMOST from simulation. $\rm N_{\rm obs}$ denotes the number of detected RRab stars with blueshifted hydrogen emission in LAMOST in our survey. Here $\rm N_{\rm theo} = 371.79$, $\rm N_{\rm obs} = 70$, $P(O)$ $\approx$ 18.83\%.

We suggest that $P(O)$ is underestimated. First, our simulation does not contain the selecting of H$\beta$ emissions due to the low resolution of the spectra, so $P(B\mid A)$ is overestimated. Another reason for the overestimation of $P(B\mid A)$ is that we used observed distribution but not the intrinsic distribution for emission flux. Moreover, in the progress of fitting we
abandon some spectra which may show shock signatures but are hard to identify or provide no valid measurements, or contain no valid H$\beta$ emissions. That is to say, more than 18.83\% of the RRab stars show relatively strong ``first apparition''.

%%%%%%%%%%%%%%%%%%%%%%%%%%%%%%%%%%%%%%%%%%
\vskip 0.5cm
%%%%%%%%%%%%%%%%%%%%%%%%%%%%%%%%%%%%%%%%%%W/D

%%%%%%%%%%%%%%%%%%%%%%%%%%%%%%%%%%%%%%%%%%

\section{Conclusions}\label{section:conclusion}

In this work, we develop a hand-crafted 1D pattern recognition searching
algorithm and apply it to a large dataset of single-epoch spectra of RR Lyrae
stars, in order to fetch out the ``first apparitions''. Through this survey, we
found 33 RRab stars and 10 RRc stars in SDSS, 70 RRab stars, 10 RRc stars, 3
RRd stars and 1 Blazhko type RR Lyrae star in LAMOST. Based on the searching
results, we set up the first population study of the RR Lyrae variables showing
hypersonic shock waves.

We build up the largest database of blueshifted hydrogen emission in RR Lyrae
stars. The features of the ``first apparitions'' are fitted by two
$S\acute{e}rsic$ profiles. We provide the redshift $z_{\rm e1,\alpha}$ and radial velocity in the stellar rest frame $V_{\rm
e1,\alpha}$, normalized flux Flux$_{\rm e1,\alpha}$ and full width at half
maximum of the emission and absorption FWHM$_{\rm e1,\alpha}$ in the stellar
rest frame of the blueshifted hydrogen emission.
The distribution of measurements for different types are compared. We provide
detailed analysis for the light curve of the Blazhko type RR Lyrae star with
ZTF DR5. We characterize this Blazhko type RR Lyrae star as an RRab star with
strong Blazhko modulations with a possible Blazhko period
$P_B\approx111.111{\rm d}$.

Finally, we set up two simulations for RRab variables in LAMOST. As for the first one, we apply our algorithm on mock spectra to test the performance of our pipeline and to check the influence of the selecting criteria. The other simulation is based on the real observational strategy to investigate the occurrence of the blueshifted hydrogen emission in RRab variables in LAMOST. The result suggests that more than 18.83\% of the RRab stars exhibit relatively strong ``first apparition''. The nature of RR Lyrae variables will be more and more clear with
enormous volume of upcoming observational data.

\vskip 1cm

\section*{Acknowledgements}

The suggestions and comments by the anonymous referee are gratefully acknowledged.  We thank the help from Dr. Hao-Tong Zhang, Dr. Zhong-Rui Bai, and Dr. Jian-Jun
Chen for getting the single-epoch spectra from LAMOST. We acknowledge
discussion with Dr. Anupam Bhardwaj. Xiao-Wei Duan acknowledges research support from the
Cultivation Project for LAMOST Scientific Payoff and Research Achievement of
CAMS-CAS and the Peking University
President Scholarship. Li-Cai Deng thanks research support from the National Science
Foundation of China through grants 11633005. Xiao-Dian Chen also thanks
support from the National Natural Science Foundation of China through grant
11903045. Hua-Wei Zhang thanks research support from the
National Natural Science Foundation of China (NSFC) under No. 11973001 and
National Key R$\&$D Program of China No. 2019YFA0405504. We acknowledge Mark Taylor
for the TOPCAT software. 
SDSS-III (\href{http://www.sdss3.org/}{http://www.sdss3.org/}) was funded by the Alfred P. Sloan Foundation, the
Participating Institutions, the National Science Foundation, and the U.S.
Department of Energy Office of Science.  
SDSS-III is managed by the Astrophysical Research Consortium for the Participating Institutions of the SDSS-III Collaboration including the University of Arizona, the Brazilian Participation Group, Brookhaven National Laboratory, Carnegie Mellon University, University of Florida, the French Participation Group, the German Participation Group, Harvard University, the Instituto de Astrofisica de Canarias, the Michigan State/Notre Dame/JINA Participation Group, Johns Hopkins University, Lawrence Berkeley National Laboratory, Max Planck Institute for Astrophysics, Max Planck Institute for Extraterrestrial Physics, New Mexico State University, New York University, Ohio State University, Pennsylvania State University, University of Portsmouth, Princeton University, the Spanish Participation Group, University of Tokyo, University of Utah, Vanderbilt University, University of Virginia, University of Washington, and Yale University. 
The Large Sky Area Multi-Object Fiber Spectroscopic
Telescope (Guoshoujing Telescope) is a National Major Scientific Project built by the Chinese
Academy of Sciences, funded by the National
Development and Reform Commission. LAMOST is operated and managed by the
National Astronomical Observatories, Chinese Academy of Sciences. 
The Catalina Sky Survey survey is supported by the National Aeronautics and Space Administration
under Grant No. NNG05GF22G.  The CRTS survey is supported by the
U.S.~National Science Foundation under grants AST-0909182 and AST-1313422. 
Data Processing and Analysis Consortium of {Gaia} (\href{https://archives.esac.esa.int/gaia}{https://archives.esac.esa.int/gaia}) was funded by
national institutions. 
The Wide-field Infrared Survey
Explorer (WISE) is supported by the National Aeronautics and Space Administration.
The All-Sky Automated Survey for Supernovae (ASAS-SN) group are supported by Gordon and Betty Moore Foundation 5-year grant GBMF5490, and NSF Grants AST-151592 and AST-1908570. The construction and
operations of ATLAS are supported by grants 80NSSC18K0284 and 80NSSC18K1575 under NEOO.
Based on observations obtained with the Samuel Oschin 48-inch Telescope at the Palomar Observatory as part of the Zwicky Transient Facility project. ZTF is supported by the National Science Foundation under Grant No. AST-1440341 and a collaboration including Caltech, IPAC, the Weizmann Institute for Science, the Oskar Klein Center at Stockholm University, the University of Maryland, the University of Washington, Deutsches Elektronen-Synchrotron and Humboldt University, Los Alamos National Laboratories, the TANGO Consortium of Taiwan, the University of Wisconsin at Milwaukee, and Lawrence Berkeley National Laboratories. Operations are conducted by COO, IPAC, and UW.

\vspace{5mm} 
\software{NumPy \citep{NumPyoliphant2006guide,Harris2020Natur.585..357H}, SciPy
\citep{Virtanen2020SciPy-NMeth}, AstroPy
\citep{Astropyprice2018astropy}, Matplotlib
\citep{Matplotlibhunter2007matplotlib},
Scikit-learn \citep{Scikitpedregosa2011scikit}, 
PyAstronomy \citep{Czesla2019ascl.soft06010C}, TOPCAT
\citep{Taylor2005ASPC..347...29T}}

\bibliographystyle{aasjournal}  %mnras_mwilliams}
\bibliography{mainnewApJDXW}

\end{document}